\documentclass{article}
\usepackage[utf8]{inputenc}
\usepackage{amsfonts}
\usepackage{amsmath}
\usepackage{amssymb}
\usepackage{graphicx}
\usepackage[numbers,compress,square]{natbib}

\renewcommand{\vec}[1]{\boldsymbol{#1}}

\newcommand{\eq}[1]{\begin{equation}#1\end{equation}}

\newcommand{\avec}[1]{\left(\begin{array}{c} #1 \end{array}\right) }

\renewcommand{\exp}[1]{{\rm e}^{#1}}
\newcommand*\Laplace{\mathop{}\!\mathbin\bigtriangleup}

\title{Estimation of functional diversity and species traits from ecological monitoring data}

\author{Alexey Ryabov$^{1,2,4}$, Bernd Blasius$^{1,2}$, Helmut Hillebrand$^{1,2,3}$,\\  Irina Olenina$^{5,6}$ and Thilo Gross$^{1,2,3,*}$}

\date{\tiny
$^1$ HIFMB, Helmholtz Institute for Functional Marine Biodiversity, Oldenburg, Germany\\
$^2$ Univeristy of Oldenburg, Institute for Chemistry and Biology of the Marine Environment, Oldenburg, Germany\\
$^3$ Alfred-Wegener Institute, Helmholtz Center for Marine and Polar Research, Bremerhaven, Germany\\
$^4$ TU Dresden, Institute for Forest Biometrics and Forest Systems Analysis, Tharandt, Germany\\
$^5$ Klaipedos Universitetas, Marine Research Institute, Klaipeda, Lithuania\\
$^6$ Lithuanian Environmental Protection Agency, Klaipeda, Lithuania\\
$^*$ Corresponding author: thilo2gross\@gmail.com
}

\graphicspath{{mainfigs/},{figs/}}

\begin{document}

\maketitle

\begin{abstract}
\noindent{}The twin crises of climate change and biodiversity loss define a strong need for functional diversity monitoring. While the availability of high-quality ecological monitoring data is increasing, the quantification of functional diversity so far requires the identification of species traits, for which data is harder to obtain. However, the traits that are relevant for the ecological function of a species also shape its performance in the environment and hence should be reflected indirectly in its spatio-temporal distribution. Thus it may be possible to reconstruct these traits from a sufficiently extensive monitoring dataset. Here we use diffusion maps, a deterministic and de-facto parameter-free analysis method, to reconstruct a proxy representation of the species' traits directly from monitoring data and use it to estimate functional diversity. We demonstrate this approach both with simulated data and real-world phytoplankton monitoring data from the Baltic sea. We anticipate that wider application of this approach to existing data could greatly advance the analysis of changes in functional biodiversity.
\end{abstract}

\section{Introduction}
Recent assessments have documented the ongoing precipitous loss of global biodiversity~\cite{cardinale2012biodiversity,mace2018aiming,ceballos2017biological,ipbes2019full}. More complex responses are observed on the regional scale, where stressors can lead to a transient increase in diversity~\cite{blowes2019geography,antao2020temperaturerelated}. Meanwhile our understanding of the complex dynamical interplay of dispersal, extinctions and speciation that has created earth's biological diversity and drives current dynamics is still woefully incomplete~\cite{gross2020modern}. Hence the scale of the unfolding crisis, the intricacy of the dynamics involved, but also the gaps in our understanding highlight the need for large-scale biodiversity monitoring.

Even quantifying biodiversity loss still poses challenges. It has been argued that, for simplicity, global policy goals should be phrased in terms of the number of extinctions~\cite{rounsevell2020biodiversity}. Similarly to climate goals, quantified by temperature-increase, the number of extinctions has the benefit of being easily communicable. However, unlike climate change, where many detrimental effects are directly triggered by rising temperatures, extinction numbers are a poor indicator of biodiversity loss, where a major concern is the loss of biological functions~\cite{tilman2001functional}.     

On a fundamental level biodiversity can be conceptualized as the genetic variation of forms, but due to the complexity of biological life the genetic makeup is only a weak indicator of function~\cite{weithoff2014comparing,ashkaanBacteria}. Hence, for the assessment of ecosystem functioning, service provision, sustainability and quantification of responses to stressors robust measures of functional diversity are needed.  

The need to understand functional diversity has been frequently highlighted \cite{tilman2001functional,petchey2002functional,hillebrand2008consequences,macdougall2013diversity,hillebrand2017biodiversity}. Common measures, such as the Rao index~\cite{rao1982diversity,ricotta2011rao}, compute functional diversity from pairwise functional distances between species. To compute such distances, researchers identify traits of the species under consideration and then compute functional diversity from distances in trait space, e.g.~\cite{walker1999plant,ricotta2011rao,mcwilliam2020deficits,nunnally2020traitbased,ramond2021phytoplankton}. 
As there's no universal definition of what constitutes a trait, it's useful to distinguish been physiological characteristics directly identified from observation (here \emph{o-traits}), and inferred traits that are inferred from data to approximate fundamental niche axis in the system \emph{i-traits}. 

Trait-based approaches provide good estimates of functional diversity, but require the researcher to quantify the trait space of all organisms considered. The decision which o-traits are relevant functional characteristics is made based on the researcher's experience and is dependent on the group of species and functions under consideration. Some traits may be difficult to measure and their values may be dependent on environmental conditions~\cite{thomas2017temperaturenutrient} and hence are context-dependent. In practice, these constraints mean that trait-based quantification of diversity is presently constrained to comparatively small groups of similar well-studied organism and suffers from limited data availability.  

In comparison to the manual determination of trait values, it's generally easier to quantify properties such as species identity, biomass and/or abundance. Long-term ecological research programs have accumulated a treasure trove of monitoring data, recording this information with individual data sets spanning multiple decades and capturing dozens or hundreds of species.  
It's therefore attractive to infer trait vales from such data sets. For example~\cite{mutshinda2017phytoplankton} used Bayesian model fitting to infer four values of o-traits from long-term timeseries. A natural next step is to use data analysis approaches to not only infer trait values, but to also construct the i-trait axes directly form data.

Here we propose an approach for the analysis of monitoring datasets that record the abundance or biomass of species observed in a set of samples.
We use diffusion maps~\cite{coifman2005geometric,moon2019visualizing,barter2019manifold}, a manifold learning method, to construct an i-trait space directly from these datasets. In contrast to previous work~\cite{mutshinda2017phytoplankton} this approach does not require a model or a list of known o-traits and isn't limited to time series data. Instead, the diffusion map identifies both the i-trait axes and trait values solely from species biomass in samples. The functional diversity can then be computed from the pairwise distances in the i-trait space. We test this approach with a simulated data set from a mathematical model, before applying it to quantify functional diversity of phytoplankton communities in a monitoring data set from the Baltic Sea~\cite{gasiunaite2005seasonality}. Our results show that the proposed method can reveal biologically meaningful trait information and allows for the robust and unambiguous quantification of functional diversity from monitoring data. In the dataset analyzed here, it reveals an increase in functional diversity with time that is significantly more pronounced at the coastal stations.   

It is interesting to note that the data-driven i-trait approach used here is diametrically opposite to traditional ecological thinking. Many classic works observed morphological features of species, conjectured their functional relevance, and then used this insight to predict spatial distribution. In the present paper we go the opposite way, using spatial co-occurrence to infer functional niches and then conjecture their potential physiological basis.    

\begin{figure}[t]
\centering
\includegraphics[width=0.8\textwidth]{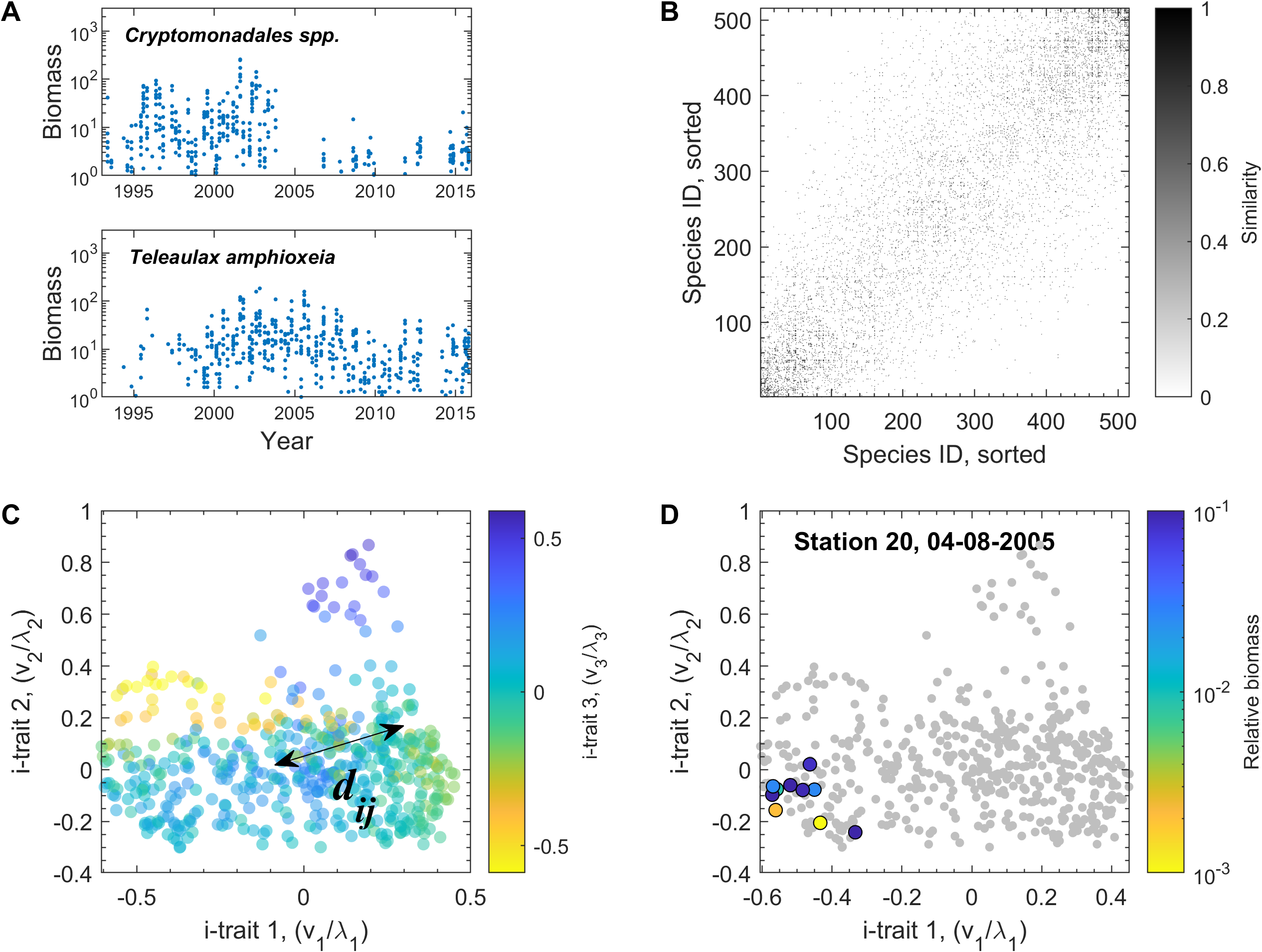}
\caption{Functional diversity estimation from a monitoring data set. {\bf A.} We use data on the biomass of 516 phytoplankton species in 730 samples collected from 1993 till 2015 from 10 stations in the Baltic sea \cite{gasiunaite2005seasonality} (two species shown for illustration). {\bf B.} We then compute the pairwise similarity between species from correlation between species abundances over the set of samples. {\bf C.} From the similarities the i-trait space of the species (dots) is constructed using diffusion maps. In this space the pairwise functional dissimilarity is quantified by the diffusion distance $d_{ij}$. {\bf D.} Once the distances between species have been determined the diversity in a specific sample can be quantified by applying the Rao's index to the species present (highlighted dots). In the sample shown most of the biomass is concentrated in a small area of trait space, leading to a comparatively low Rao index.         
\label{fig1construction}}
\end{figure}

\section{Trait space inference from monitoring data}
Quantifying differences between dissimilar objects poses a fundamental challenge: Whereas we may be able to compare two songs or two paintings, it is much harder to quantify how dissimilar a certain painting is from a certain song. The same challenge is encountered in assessments of functional diversity, where it is essential to quantify how dissimilar pairs of (potentially very different) species are.   
To circumvent this problem, the diffusion map \cite{coifman2005geometric,moon2019visualizing,barter2019manifold} builds on the idea that the dissimilarity between pairs of objects can be robustly quantified if they are sufficiently similar. By finding all such short-distance comparisons that can be made in the dataset we obtain a set of ``trusted'' links between objects. 

To apply this approach to functional diversity estimation we quantify the similarities between species based solely on their abundance in monitoring samples. Our primarily notion of similarity is the Spearman correlation~\cite{spearman1904proof} between pairs of species across samples in the dataset, which provides an indicator of co-occurrence of species. 

We follow \cite{barter2019manifold} and consider a comparison between two species as a trusted link if it ranks in the top-10 most similar comparisons for at least one of the two species. The trusted similarities are stored in a similarity matrix $\bf S$, while all others are set to zero. The result is a network of trusted links that spans the entire set of species while containing only relatively short-ranged and hence relatively accurate comparisons.

Once trusted links have been identified, we quantify the dissimilarity between species by their distance in the network of trusted links. Specifically, diffusion maps use the notion of diffusion distance \cite{coifman2005geometric}, which takes all possible paths between network nodes into account. 
We use a variant of diffusion distance $d_{ij}$, which can be computed efficiently from the set of eigenvectors and eigenvalues of a Laplacian matrix describing the network, see appendix for details. 
The result is a computationally efficient method (Fig.~\ref{fig1construction}) that produces deterministic results, and where our choice of trusting ten neighbors is the only tuneable parameter.

The Laplacian eigenvectors that are identified in the process are also of interest for a different reason: The $n$th eigenvector $\vec{v_n}$ contains one element corresponding to each of the species, which is related to the $n$th i-trait for that species. The corresponding eigenvalue $\lambda_n$ is inversely proportional to the relative importance of this $n$th trait axis (see appendix). Hence the rescaled vector $\vec{v_n} / \lambda_n$ specifies the properly scaled value of the $n$th i-trait of all species in an effective i-trait space (Fig.~\ref{fig1construction}C). We show below that these i-traits align well with ecological intuition.

Once the i-trait space has been constructed we consider individual samples from the monitoring dataset (Fig.~\ref{fig1construction}D). Building on the inter-species diffusion distances in the i-trait space, we quantify the functional diversity in the sample using Rao's quadratic entropy~\cite{rao1982diversity}. The method can thus quantify functional biodiversity and, to some extent, place the species into a biologically meaningful trait space. 

\section{Validation with model data}
\begin{figure}[htb]
\centering
\includegraphics[width=\linewidth]{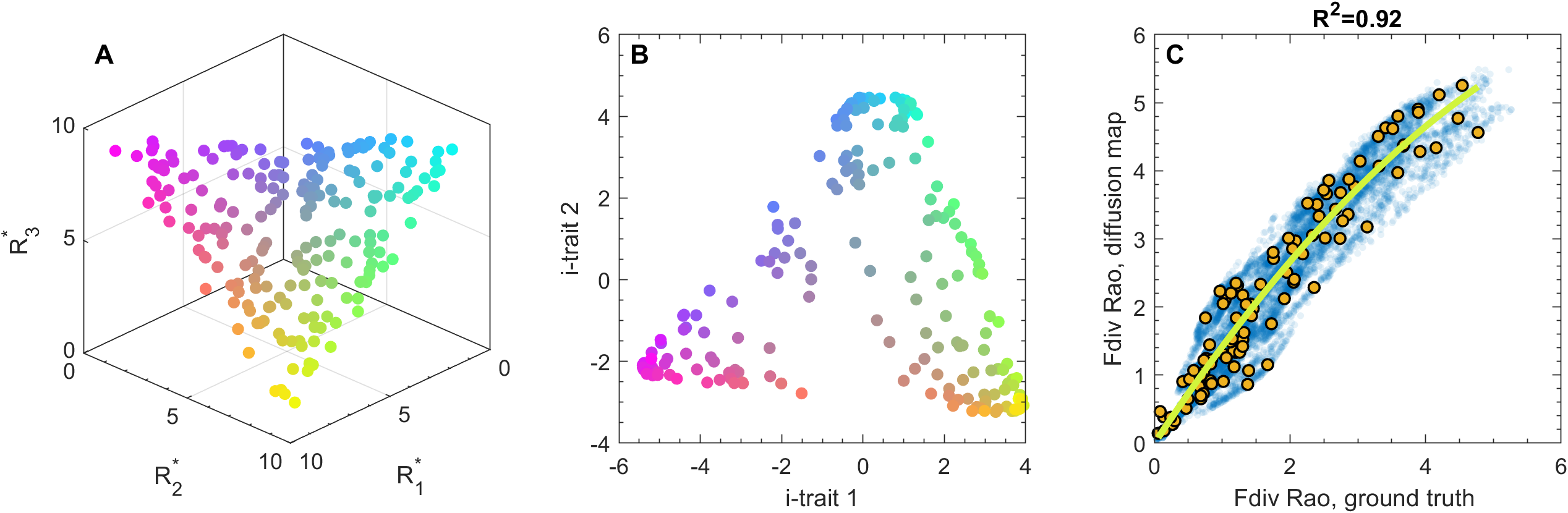}
\caption{Numerical validation of the proposed method. {\bf (A)} We numerically generate randomly distributed traits of 200 species (colored dots) that fill a triangle in a trait space spanned by three resource-requirement ($R^*$) parameters.
Color indicates the resource ratio preferred by a species. {\bf (B)}
Inferred i-trait space generated by diffusion mapping simulated biomass data. The reconstruction  identifies traits that span a space that is qualitatively similar to the ground-truth traits. 
Colors are the same as in A, illustrating that neighborhood relationships are mostly reconstructed correctly.  
{\bf (C)} Rao's functional diversity calculated from diffusion distances in the reconstructed trait space correlates strongly with the numerical ground truth (based on $R^*$ values). Indicated are local diversity (blue dots) in individual patches and regional diversity in meta-community (yellow dots). The R$^2$ value for regional diversity is 0.92, relative to a cubic regression (green line). These results show that the proposed method can be used to identify traits and robustly estimate functional diversity based on monitoring data.}
\label{fig2validation}
\end{figure}

To test the proposed method we generate synthetic data by simulating a meta-community model \emph{in silico}. We consider a community of 200 primary producer species limited by three essential resources. Each species is characterized by a set of minimal resource requirements ($R^*$ values) reflecting the species ability to sequester the corresponding resource. We can thus envision the species as points in a three-dimensional trait space (Fig.~\ref{fig2validation}A), where the $R^*$ values correspond to the traits. Specifically, we randomly draw the $R^*$ values such that they fill a triangular surface, modeling the existence trade-offs such that a greater ability in sequestering one resource is compensated by a lesser ability in sequestering others.  

We adapt a model from~\cite{hodapp2016environmental} to simulate the population dynamics of species in 800 meta-communities, each of which consists of a square lattice of 120 discrete patches arranged in a $10 \times 12$ grid. The individual patches are characterized by random values describing supply of three resources, mimicking real world spatial heterogeneity and facilitating coexistence of model species. The biomass density of each species in each patch changes dynamically according to an equation capturing local growth and mortality as well as dispersal to and from neighboring patches (see appendix). The result is a spacial meta-community in which all species persisted over the duration of simulation runs. 

As a first test, we consider the distribution of species in the space spanned by the two most important i-traits, found by diffusion mapping the simulated biomass samples. While the i-trait space is slightly deformed in comparison to the ground truth, it retains key characteristics (\emph{cf.} Fig.~\ref{fig2validation}A,B). In the i-trait space the species still form a triangular shape on a two-dimensional surface. This result gives us confidence that the diffusion map should also be capable of inferring meaningful trait-spaces from large real-world datasets.     

The main purpose of the proposed method is the estimations of functional diversity. To test whether a trait space that has been infered from one dataset can also be used to estimate the diversity in new samples we ran 100 additional simulations and estimated the functional diversities both for the ground truth, given by the known $R^*$ values, and for the i-traits using the existing map. This was done both for the entire meta-community (mimicking regional diversity) and within each patch (local diversity). A comparison of the resulting Rao indices (Fig.~\ref{fig2validation}C) shows a strong correlation ($R^2 = 0.92$) between the ground truth and the reconstructed values of regional diversity.     

We also explored how limited data availability and different distance metrics impacts the accuracy of reconstruction (see appendix). The power of the diffusion map hinges on our ability to construct a spanning network of trusted comparisons between the samples. If the underlying trait space is large or the number of species is small then we are forced to trust comparisons between comparatively dissimilar species, and the quality of the reconstruction degrades. By contrast, a larger number of observations reduces the noise in individual comparisons and improves the quality reconstruction (Figs.~8,10).

\begin{figure}[htb]
\centering
\includegraphics[width=\linewidth]{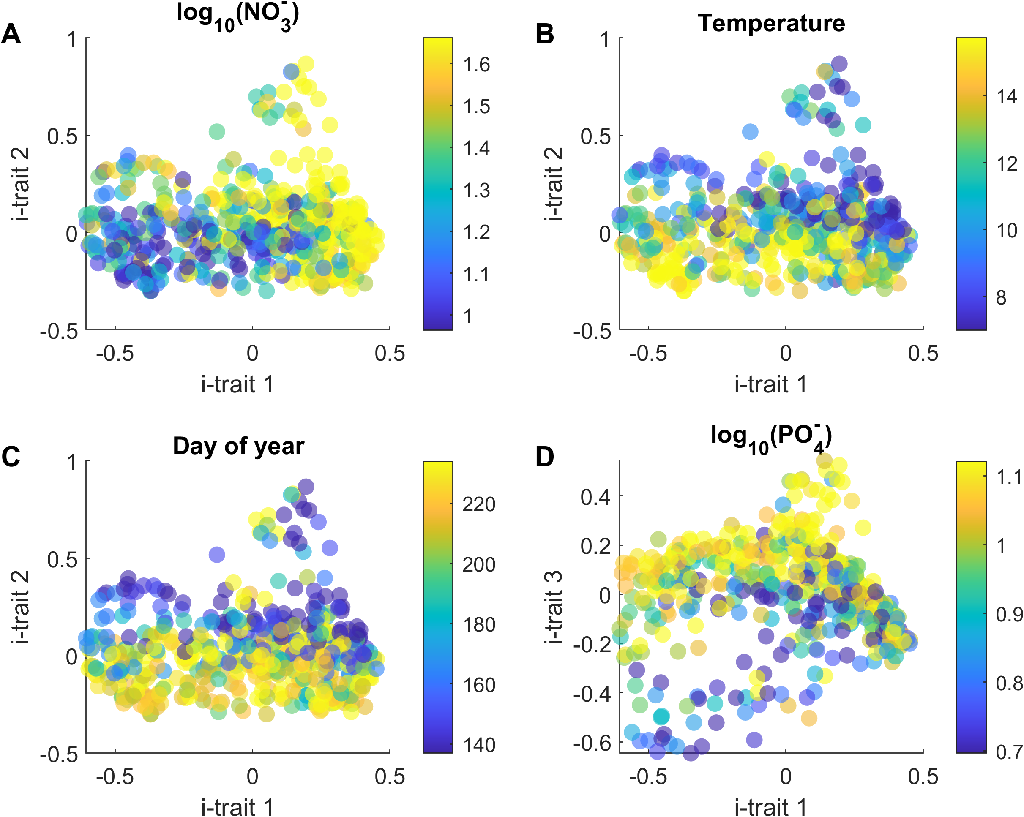}
\caption{Inferred traits from the monitoring dataset. Shown are species (dots) projected onto the space spanned by the most important i-traits. Color-coded are environmental conditions under which the species were observed with high relative abundance (see text).
The first i-trait aligns well with NO$_3^-$ concentration separating species by their nitrogen requirements. The water temperature (B) and day of the year (C) align with the second trait, separating the early from the late species. The PO$_4^-$ concentration is closely aligned with the third reconstructed trait (D).  \label{fig3traits}}
\end{figure}

In summary, results from the numerical experiments show that, given a sufficient volume of data, the diffusion map can infer the trait space from a monitoring dataset. Moreover, independently of the interpretation of the trait space, it can be used to robustly quantify the dissimilarity between species, which allows to infer functional diversity from monitoring data.  

\section{Analysis of Baltic Sea phytoplankton species}
We now turn back to the phytoplankton monitoring dataset (Fig.~1). The data was collected in the Lithuanian coastal area of the Baltic Sea and spans a period from May to November for 23 years (1993-2015). In total it contains $730$ samples of the biomasses of $516$ species measured at different times and stations (see Fig.~\ref{fig1construction}A for examples). We analyze the Baltic data using the same procedure that we applied to the simulation results. A projection of the trait space using the most important i-traits is shown in Fig.~\ref{fig1construction}C. 

Diffusion mapping does not provide a biological interpretation of the inferred traits. However, we can uncover such an interpretation for at least some of the traits by analyzing additional data. Here we discuss in particular four environmental variables that were recorded during sampling (day of year, water temperature, NO$_3^-$ concentration and PO$_4^-$). For each of the species we calculate the mean environmental conditions at which it was observed. This is done by computing a weighted average of each environmental parameter, where the biomass of the species under consideration is used as the statistical weight of the sample. Color coding the species in the reconstructed trait space (Fig.~\ref{fig3traits}) shows that the first i-trait aligns well with NO$_3^-$ concentrations (Spearman correlation, $r_S=0.55$). We conclude that this trait represents adaptation to different levels of nutrient availability.

We note that the alignment of the i-trait with NO$_3^-$ is a purely statistical finiding, which doesn't necessarily imply any causal link. The interpretation as adaptation to different levels of nutrient availability must therefore be treated as a working hypothesis (cf.~Fig.~16).  
Similarly i-trait 2 closely aligns with the temperature ($r_S=0.50$) and day of year ($r_S=0.43$), suggesting that this trait represents the growth strategy, separating early from late species. 
The third i-trait correlates with the PO$_4^-$ concentrations ($r_S=0.45$). 

The number of i-trait axes equals the dimensionality of the input data, i.e.~the number of species. Although each i-trait contains less information than the previous one, projections of the trait space on different trait axes give additional insights into the distribution of species ecotypes (see appendix). 

\section{Diversity gains on the Lithuanian Coast}
Once the i-trait space has been constructed it can be used to quantify the functional diversity. We first use the i-traits from the analysis of the whole dataset to compute the diffusion distances between all pairs of species. For each sample we then use the distances between the species in the i-trait space to compute the Rao index.

\begin{figure}[ht]
 \centering
 \includegraphics[width=0.95\linewidth]{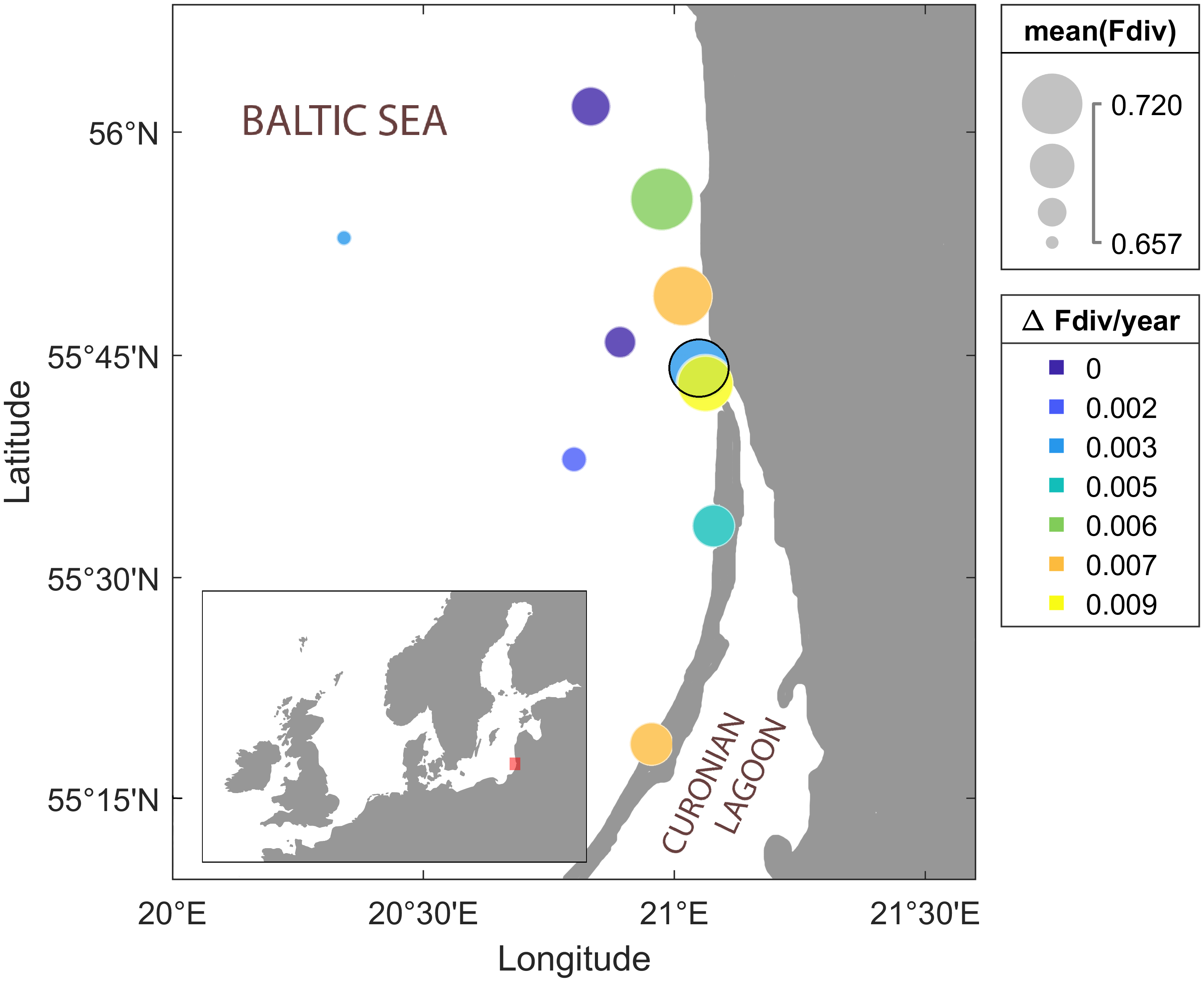}  
    \caption{Phytoplankton diversity on the Lithuanian coast. We observe an increase in functional diversity over the measurement period at all of the 10 stations included in the dataset (circles). The station located at the exit of the Curonian Lagoon is marked with a black circle. The fastest increase (warmer colors) is found at some of the coastal stations. The coastal stations are also the most diverse in average (larger diameter). }
    \centering
    \label{fig4trends}
\end{figure}

The estimated day-to-day functional diversities are relatively noisy, likely due to intrinsic fluctuations in the system. However, when considered over the whole period, there is a significant biodiversity gain at all stations, Fig.~\ref{fig4trends}. 
This gain is most pronounced at the coastal stations where the functional diversity is also the highest in the later years.

The local increase in functional diversity is consistent with previous findings and predictions. For example~\cite{blowes2019geography,hillebrand2017biodiversity,hillebrand2020integrative} observe comparable increases in species richness. A key mechanism in this context is that species extinction events triggered by environmental change take longer to manifest than corresponding invasions~\cite{hillebrand2017biodiversity}. Hence environmental disturbances are likely to trigger transient increases in biodiversity on a trajectory that eventually leads to diversity loss when either longer times or larger geographical scales are considered.  

In the present case the difference between coastal and offshore stations provides strong evidence supporting the hypothesis of an invasion-triggered transient increase. The coastal stations are close to the freshwater communities of the Curonian Lagoon. An increased influence of the freshwater species due to changing environmental conditions could easily explain the observed diversity trends.  
We note that the only coastal station that did not experience any increase of functional diversity, is located directly at the exit of the Curonian Lagoon 
(black circle in Fig.~\ref{fig4trends}) and
hence has always been strongly influenced by the freshwater communities (see appendix). 

The estimated diversity is consistent with expectations based on species composition. The low functional diversity in the spring samples 
 of early 1993 and 2000 (see S14,S15) coincides with the dominance of dinoflagellates, mainly {\it Peridiniella catenata}, whose numbers were over 50\% of the total phytoplankton abundance, of up to 96\% by biomass.
During the period of increased functional diversity 
in spring samples from 1994 to 1999, {\it Peridiniella catenata} was found in small numbers and the community was dominated by 3-5 species constituting together more than 50\% of the total abundance. During this time, the number of non-dominant species with relative abundance less than 10\% also increased. 


\section{Conclusions}
In this paper we proposed a method by which functional trait axes and values can be inferred from monitoring data. This enables a robust estimation of functional diversity within the system, based solely on species abundances or biomasses. 

We demonstrated the method using simulated data and a phytoplankton monitoring dataset from the Baltic sea. The analysis of the real world data identified adaptation for early/late growth, high/low nitrogen levels and high/low phosphorus levels as the most important functional trait axes. It also showed a local increase in functional diversity that is comparable to previous observations in other systems~\cite{blowes2019geography,hillebrand2017biodiversity,hillebrand2020integrative}. In the present analysis, the increase is most pronounced at coastal stations and can be linked to increasing influence from a nearby freshwater community. Hence our results provide additional evidence for the hypothesis that changing environmental parameters may lead to a transient increase in local diversity, that might ultimately lead to biodiversity loss on longer and larger scales~\cite{blowes2019geography,antao2020temperaturerelated}. 

We note particularly that the proposed method does not require manual identification of relevant traits. It thus provides an objective and procedurally-grounded definition of functional diversity that is transferable between different systems and sets of species. We expect that this will also be useful in the analysis of other datasets, particularly those which contain a large number of species.  

In principle the proposed method could also be applied to study bacterial diversity. However in bacteria the relevant traits are thought to be more closely related to their genetic makeup than in eukaryotes\cite{ashkaanBacteria}. Hence it is sensible to take the available genomic information into account when constructing diffusion maps of bacteria.  

One of the main motivations for the present work was our desire to eventually gain a deeper understanding of the dynamics of complex metacommunities. The trait axes that the proposed method infers can be interpreted as dynamical variables at the community level. Given sufficient data, we may eventually be able to infer also the equations that capture the dynamics directly at this level.  

The accuracy of the results should increase with the number of observations, which makes it attractive to combine multiple, and perhaps even all available, datasets for a given group of species. Different monitoring datasets should be relatively easy to fuse at this level, because only comparisons within samples from the same dataset need to be made. In the future, a diffusion map based on a large-scale aggregation of many different monitoring datasets could effectively provide a functional diversity standard that can be used to quickly map the functional diversity of samples on a fixed scale.               

\section*{Acknowledgements}
This work was funded by the Ministry for Science and Culture of Lower Saxony (HIFMB project) and the Volkswagen Foundation  (grant number ZN3285) and the German Research Foundation (DFG) in the Research Unit DynaCom Grant Number
FOR 2716. AR was partly supported by the Federal Ministry of Education and Research BMBF Germany Project PEKRIS II (03F0828)

\pagebreak

\appendix

\section{Computational Experiment}
In this section we describe the detailed procedure for the simulation of the metacommunity model, see \cite{hodapp2016environmental} for the model analysis. We model spatial competition of $n=200$ species for three limiting resources, heterogeneously distributed across a two-dimensional grid comprising of $10\times 12$ diffusively coupled cells. The resource availability in each cell $(x,y)$ is characterized by the equilibrium concentration of the three resources $(S_{xy}^{1}$, $S_{xy}^{2}$, $S_{xy}^{3})$ in the absence of consumers.

To obtain samples from environments with varying heterogeneity, we simulated 800 grids with different levels of resource variability. The regional variability of  resource $r$ in grid $j$ is constrained by the parameters $S_{\rm min}^{jr} < S_{\rm max}^{jr}$, drawn randomly from a uniform distribution in the range from 1 to 39 resource units. The local supply, $S_{xy}^{jr}$, of resource $r$ in cell $(x,y)$ of grid $j$ is then drawn randomly from a uniform distribution in the range $[S_{min}^{jr},S_{max}^{jr}]$ in the respective resource grid.

Following the mechanistic theory of resource competition, we use the $R_{ir}^*$ values, i.e., the concentration of resource $r$ at which species $i$ can still persist, as basic species traits \cite{tilmanresources1980}. To define the trait of species $i$, we draw the values $R_{i1}^*$ and $R_{i2}^*$ from a uniform distribution in the range $[0.5,9.5]$ and then chose the third resource parameter $R_{i3}^*$ such that the sum $\sum_{r=1...3} R_{ir}^* =19.5$ is constant. We reject parameter combinations for which $R_{i3}^*$ is greater than $9.5$. The trait values that are thus selected are uniformly distributed in a triangle in the trait space, as shown in Fig.~2A. 

We model the functional dependence of biomass growth using Monod kinetics of resource uptake and Liebig's law of minimum \cite{leon1975competition,huisman2001fundamental},
\eq{g_i (\vec{R})=g_{\rm max} \min_{r=1\ldots{}3}\frac{R_r}{K_{ir}+R_r},}
where $g_{\rm max}$ is the maximum growth rate, $R_r$ is the local concentration of resource $r$, and $K_{ir}$ is the half-saturation constant of growth of species $i$ limited by resource $r$. We assume that the maximum growth rate $g_{\rm max}=1$ and mortality rate $m=0.25$ are identical for all species, so that the dimensionality of the species trait space is defined by the R* values only.
Based on the condition $g_i (R_{ir}^*) = m$, we can express the half-saturation constants for given values $R_{ir}^*$ as
\eq{K_{ir}=R_{ir}^*\frac{ (g_{\rm max}-m)}{m}.}
The system is initialized by assigning random biomass values in $[0, 0.1]$ to each species in each grid cell. These biomass values are small compared to the typical local equilibrium biomass of persisting species. 

To describe the evolution of the system, we denote $N_{xy}^i$ as the local biomass of species $i$, and $R_{xy}^r$ the local concentration of resource $r$, in cell $(x,y)$. The biomass of species $i$ then evolves according to  
\eq{\frac{\rm d}{{\rm d}t}N_{xy}^i=\left[g_i(\vec{R_{xy}})-m\right]\cdot N_{xy}^i+\Delta N_{xy}^i,}
where the first term captures local growth and loss processes and the second term captures dispersal in terms of the discrete Laplace operator  
\eq{\Delta N_{xy}^i=\frac{ N_{x,y+1}^i+ N_{x,y-1}^i+ N_{x+1,y}^i+ N_{x-1,y}^i-4 N_{xy}^i}{h^2}, 
}
where $h$ is the lattice constant. At the grid boundaries we assume no flux boundary conditions. The dynamics of resource $r$ is given by the difference between local resource inflow and resource consumption (for simplicity, we neglect the diffusion of resources)
\eq{
\frac{\rm d}{{\rm d}t} R_{xy}^r = D(S_{xy}^r-R_{xy}^r)-\sum_{i=1}^n c_{ir} N_{xy}^i \, g_i ({\vec{R_{xy}}} )         
}
where $c_{ir}$ is the amount of resource $r$ consumed for the production of one unit of biomass of species $i$ and $D$ is the dilution rate. We assume that $c_{ir}$ is linearly related to $R^*$, $c_{ir} =0.05R^*_{ir}$ \cite{hodapp2016environmental}. This parameterization provides an additional possibility for species coexistence not only due to species diffusion across the grid but also due to local resource partitioning.  

\begin{figure}
    \centering
    \includegraphics[width=0.7\textwidth]{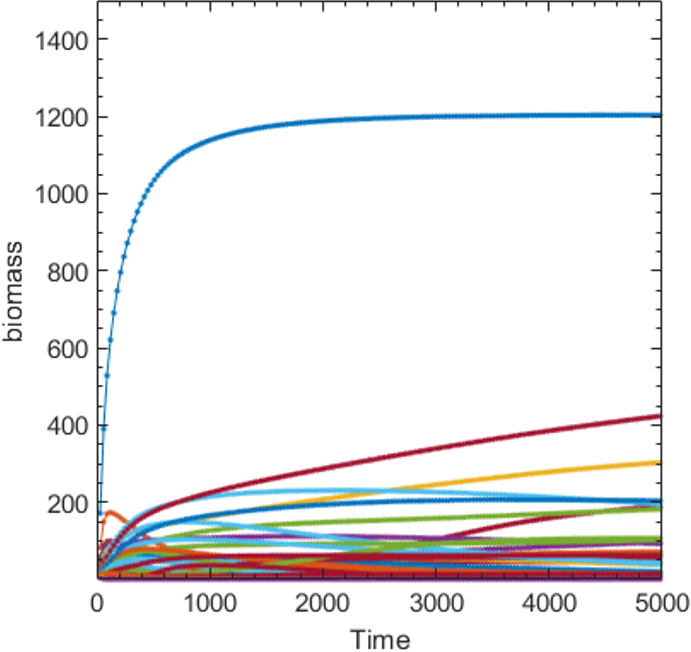}
    \caption{Typical timeseries from a simulation run. Populations approach different stationary levels, however, no competitive exclusion occurs due to the spatial distribution of resources in the metacommunity.}
    \label{figTimeserie}
\end{figure}

We integrate the system numerically for 6000 simulation days. This simulation time is sufficient for species sorting and reducing the effect of initial conditions. If the local biomass of a species drops below 0.01 at the end of the simulation, we regard this species to be locally extinct and set its local biomass to zero  (for comparison, the maximum local biomass of species is about 100). However, due to the spatial structure of the system global extinctions did not occur. An example timeserie is shown in Fig.~\ref{figTimeserie}.

The local biomass of the 200 competitors at the end of the simulation are then analyzed using the diffusion map procedure described below. 

Thus, the raw data includes $10\cdot{}12\cdot{}800=96,000$ biomass samples of each of the $200$ species. However, diffusion fluxes across cell boundaries lead to biomass correlations in neighboring cells, so that many samples obtained from the same grid are not independent. We estimate the minimal number of independent samples that we use below as the number of simulated grids.

Unless not otherwise specified above, parameter values are taken according to \cite{hodapp2016environmental}.
For numerical integration we use an explicit Runge-Kutta (4,5) algorithm (ode45 solver, MATLAB 2020).
The source code for the simulation and data analysis is publicly available at \cite{code}.


\section{The why and how of geometric diffusion}
A central challenge in the analysis of high-dimensional data is to construct faithful metrics for the dissimilarity of distant points in data space. The idea of the diffusion map is to address this challenge by the construction of a network of comparisons that are sufficiently close and can hence be trusted. 

Once such a network of trusted links has been identified we can quantify the distance between dissimilar points as the distance on the network. Perhaps the most natural way in which one could do such comparisons is to consider shortest path distances: The distance between any pair of data points is defined as the length of the shortest path on the network of trusted links. 

Such a quantification in terms of shortest path distance does indeed work, and provided sufficient data yields very similar results to the diffusion map. However, shortest path distance is very susceptible to noise as it hinges on the presence or absence of specific links. Therefore this notion of distance is typically not used in manifold learning. 

To increase the robustness to noise one needs to take the length of alternative (non-shortest) paths between data points into account. For this reason Coifman \cite{coifman2005geometric} proposed the notion of diffusion distance, where the distance between points is quantified in terms of properties of random walks on the network. As a process the random walk effectively explores all paths in the network, leading to excellent robustness. 

There are some different variants of the diffusion process that each come with their unique advantages and disadvantages. The most common diffusion process studied in physics is continuous time diffusion. In this case walkers move in discrete jumps, but these jumps occur randomly at random times such that the average rate is 1 according to a Poisson process. The probability to find a walker in node $i$ at time $t$, $x_i(t)$ then changes according to the following differential equation
\eq{
\dot{x}_i = -k_ix_i + \sum_j A_{i,j} x_j,
}
where $A_{i,j}$ is the adjacency matrix of the network and $k_i=\sum_j A_{i,j}$ is the degree of the node. In our case we take this adjacency to be the similarity along trusted links $S_{i,j}$ or zero if there isn't a trusted link between a certain pair of nodes. 

The equation above can be written in vector form as 
\eq{
\dot{\vec{x}} = - {\bf K}\vec{x} + {\bf A}\vec{x}, 
}
where $\vec{x}=(x_1,\ldots,x_N)^{\rm T}$, ${\bf A}=(A_{ij})$ and 
$\bf K$ is the degree matrix defined by
\eq{
K_{i,j}= \left\{\begin{array}{l l} k_i & \quad  \mbox{for $i=j$} \\ 0 & \quad \mbox{otherwise}\end{array}\right. .
}
Defining the Laplacian ${\bf L} = {\bf K}-{\bf A}$ allows to write this equation more compactly as  
\eq{
\dot{\vec{x}}= - {\bf L}\vec{x}.
}
The sign of $\bf L$ was chosen to make the operator $\bf L$ positive semidefinite. Apart from this unfortunate choice, the equation is closely reminiscent of the diffusion equation in continuous space ($\dot{\vec{x}}=\Laplace \vec{x}$). Where the Laplace operator ($\Laplace$) has been replaced with the Laplacian matrix $\bf L$. In fact, the Laplacian matrix can be read as the discretization of the second spatial derivative $\Laplace$ on a lattice.

Since the diffusion equation is linear we can solve it by eigen-decomposition and hence 
\eq{
\vec{x}(t)=\sum c_n \vec{v_n} \exp{-\lambda_n t},
}
where the $c_n$ are the expansion coefficients of an initial state $\vec{x}(0)$ with respect to the eigenvectors $\vec{v_n}$ of $\bf L$, i.e.~${\bf L}\vec{v_n}=\lambda_n\vec{v_n}$ for all $n$ and $\sum c_n \vec{v_n}=\vec{x}(0)$.  

Laplacian matrices of connected networks are positive semi-definite matrices that have exactly one eigenvalue $\lambda_1=0$, with the corresponding eigenvector $\vec{v_1}=(1,\ldots 1)^{\rm T}$. All other eigenvalues are positive. Hence, for $t\to \infty$ all nodes will eventually approach the same state. However, during transient dynamics the occupation probabilities $x_i$ of nodes can be different. Specifically, we can write the instantaneous difference between the state of nodes $i$ and $j$ as   
\eq{
d_{i,j} = \sum c_n (v_{n,i}-v_{n,j}) \exp{-\lambda_n t}
}
We can now quantify the dissimilarity of two nodes in terms of the integral over their difference $d_{i,j}$ following a small perturbation
\eq{
D_{i,j} = \int_{t=0}^\infty \sum c_n (v_{n,i}-v_{n,j}) \exp{-\lambda_n t}  = \sum c_n \frac{v_{n,i}-v_{n,j}}{\lambda_n}.
}
We can see here that differences in the direction of the $n$-th eigenvector are scaled by the factor $1/\lambda_n$. This gives infinite weight to the first eigendirection, but this is not a problem as all $v_{1,i}-v_{1,j}=0$ for all $i,j$. We can see that the most important eigendirections are those with positive eigenvalues close to zero. This makes intuitive sense as perturbations that excite the corresponding eigenvectors will decay very slowly and hence have a long impact on the system. Nodes whose coordinates differ strongly in these important vectors will generally be located far apart from each other.  
The calculation above motivates interpreting $v_{n,i}/\lambda_n$ as the $n$-th coordinate of node $i$ in a newly defined space, which traces the manifold and is properly scaled for comparisons between nodes. 

Despite it's physical grounding the unnormalized Laplacian is rarely used in diffusion maps because it also has some disadvantages. Perhaps most importantly it is a symmetric matrix and hence its eigenvectors are always orthogonal. In a series of tests we found that this limits the eigenvector's ability to align with natural features of the data.

Common implementations of the diffusion map\cite{coifman2005geometric} build on the idea of discrete-time diffusion. In this case one considers a system where a walker moves in discrete time intervals by following one of the available links in every time step. This leads to a discrete time evolution equation of the form 
\eq{
x_i(t+1)=\sum A_{ij} \frac{x_j(t)}{k_i} 
}
where $k_j$ is the weighted degree of node $j$, i.e.~the sum of the similarities along all trusted links connecting to this node. We can write this system in the vector form 
\eq{
\vec{x}(t+1)= {\bf L'} \vec{x}(t)
}
where $L'_{i,j}=A_{i,j}/k_j$. This matrix is closely related to the physical Laplacian $\bf L$. We can compute it from $\bf L$ by dividing each row by the diagonal element, subtracting the identity matrix and inverting the sign of the remaining entries. The result is a matrix with eigenvalues in the interval $[-1,1]$, where the most important eigenvalues are now the ones close to $1$. By computing the first passage times between nodes one can rigorously motivate a new coordinate system where the $n$-th coordinate of node $i$ is now,
$v_{n,i}\lambda_n$, where $\vec{v_n}$ is again the $n$-th eigenvector \cite{coifman2005geometric,coifman2006diffusion,nadler2006diffusion}.     

One of the disadvantages of ${\bf L'}$ is that time discretization causes some discretization artifacts that can lead to unphysical behavior. For example on a network that consists of a single linked pair, the walker will be on the starting node in all odd steps and on the respectively other node in all even steps, and hence the occupation probability never equilibrates. The artifacts are less pronounced in larger, more generic networks but generally the negative eigenvalues represent such unphysical modes.

\begin{figure}
    \centering
    \includegraphics[width=0.49\textwidth]{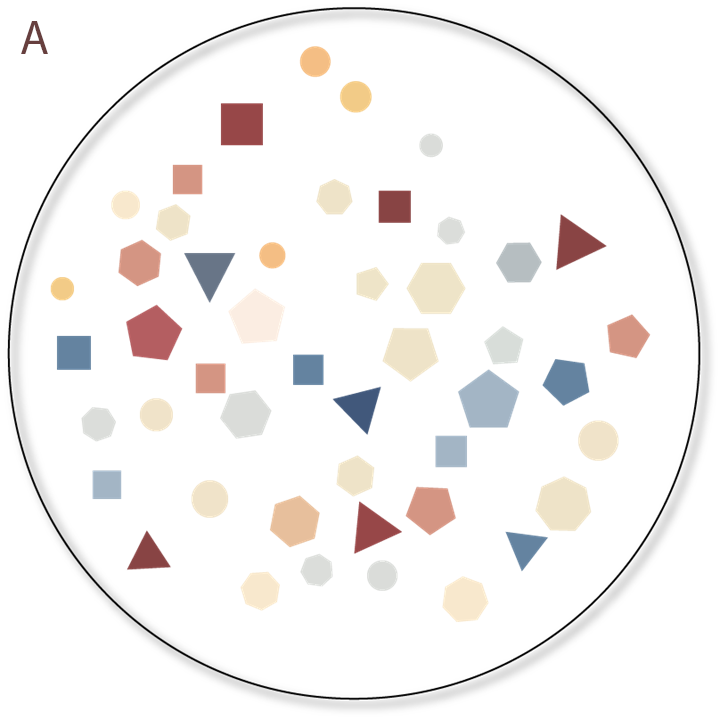}
    \includegraphics[width=0.49\textwidth]{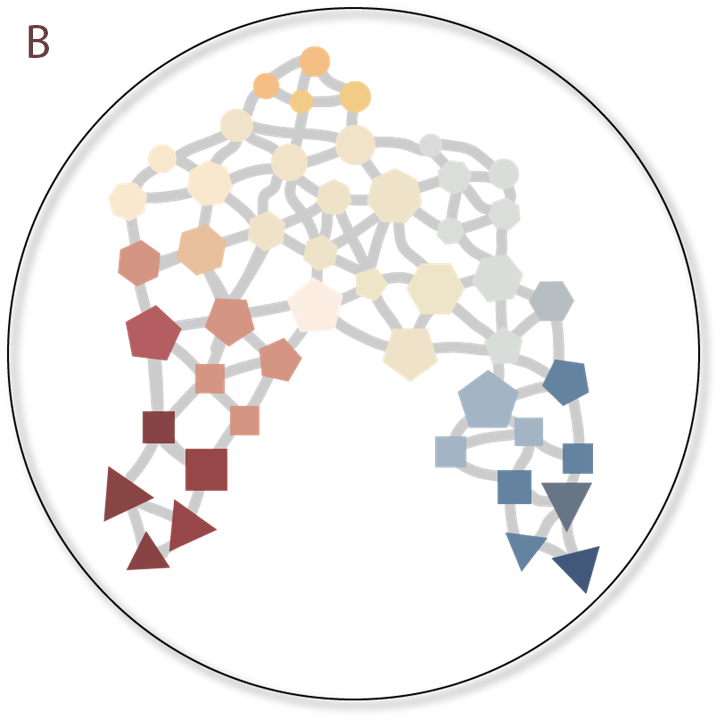}\\
    \includegraphics[width=0.49\textwidth]{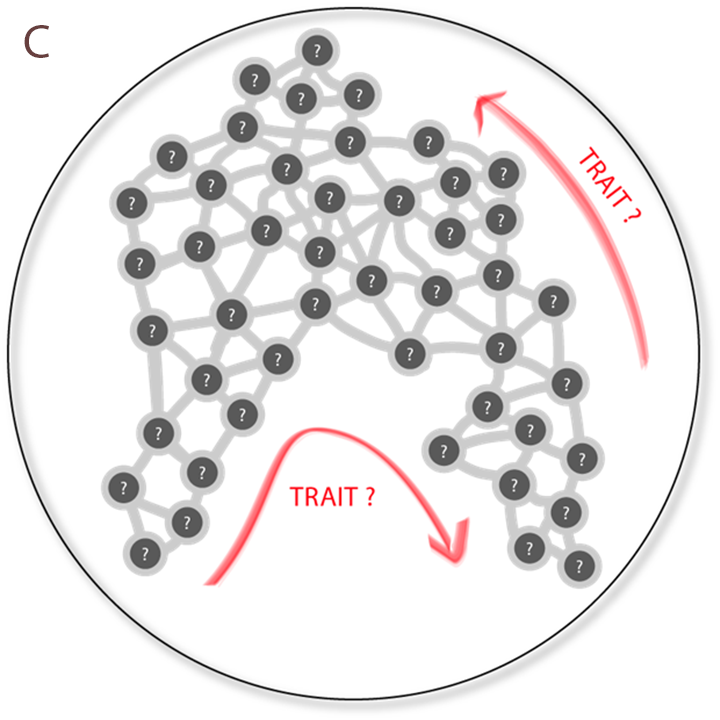}
    \includegraphics[width=0.49\textwidth]{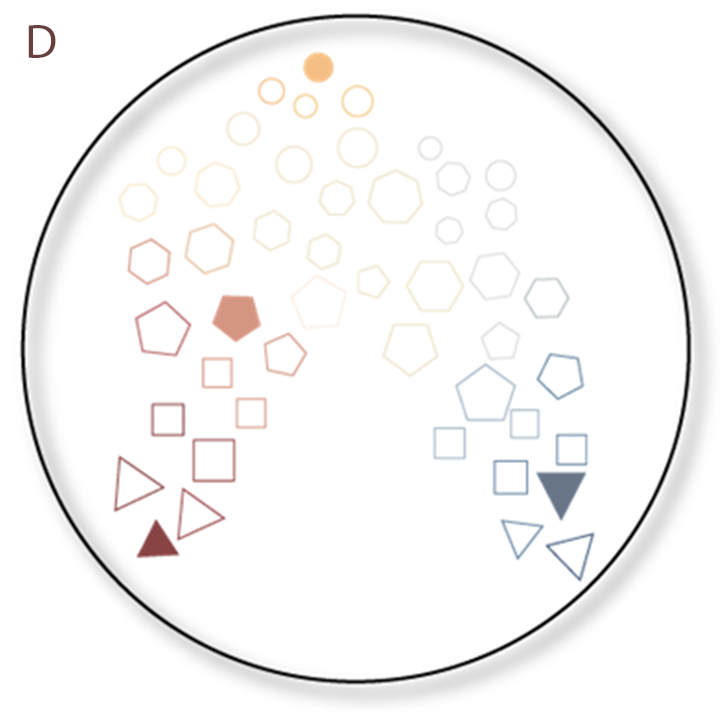}
    \caption{\textbf{Illustration of functional diversity mapping.} When dealing with large collections of dissimilar species (symbolized in A) it is useful to identify the network where the links are placed between species that are so similar that they can be faithfully compared (B). Here we construct such a network based solely on the abundance of species in samples. Although no trait information is available, traits can be inferred by identifying the major directions that span the network from the eigenvectors of a diffusion operator (C). Once the entire set of species has been assigned values of the inferred trait, long distance comparisons between species become possible, allowing us to robustly compute the functional diversity (D). In the example (D)  a set of four species covers a large part of the variation seen in the whole dataset, hence such a sample would have a high diversity despite the low number of species.}
    \label{fig:my_label}
\end{figure}

Neither ${\rm L}$ nor ${\rm L'}$ is intrinsically right or wrong as both can be advantageous or disadvantageous in certain situations. 
Mathematical work even favors a third Laplacian, the symmetrically normalized matrix
\eq{
L''_{ij} = \left\{ \begin{array}{l l} 1 & \quad \mbox{for $i=j$} \\
-A_{ij}/\sqrt{k_ik_j} & \quad \mbox{otherwise} \end{array} \right. ,
}
which cannot be motivated by common diffusion processes, but is closely related and has advantageous mathematical properties \cite{alon1986eigenvalues}. 

In a previous paper\cite{barter2019manifold} we use the row normalized Laplacian 
\eq{
L'''_{ij} = \left\{ \begin{array}{l l} 1 & \quad \mbox{for $i=j$} \\
-A_{ij}/k_{i} & \quad \mbox{otherwise} \end{array}. \right.
}
This Laplacian can be seen as a compromise between $\bf L$ and $\bf L''$. It can be derived from a physical continuous time diffusion process on a rescaled network. It can thus benefit from the wealth of physical intuition on continuous time diffusion processes. Like $\bf L$ it has a positive semidefinite spectrum with a single zero eigenvalue (in connected networks). At the same time it's eigenvectors are identical to those of ${\bf L'}$.  

While working on previous work, including \cite{ashkaanBacteria,barter2019manifold} and others, we have at several times compared diffusion maps that were obtained with different variants of the Laplacian. The results where in all cases very similar, but in our experience the matrix $\bf L'''$ typically yields the best results. We have therefore also used $\bf L'''$ as the Laplacian for the present paper. Due to it's link to the continuous time diffusion process the appropriate scaling for the new coordinates is the same as for $\bf L$ such that in new coordinates then $n$th variable of node $i$ is $v_{n,i}/\lambda_n$.

\section{Diffusion map procedure}
We now explain the procedure that was used to construct i-traits and to compute diffusion distances between species.  
\subsection{Notation} We consider $m$ samples containing a subset of a total of $n$ species. The biomass (or abundance) of species $i$ in sample $j$ is denoted by $a_j^{(i)}$. We can then collect all observations of species $i$ in the vector 
\eq{
\boldsymbol{a^{(i)}}=\avec{a^{(i)}_1\\ \vdots \\ a^{(i)}_m}
}

\subsection{Similarity matrix}  
By comparing different approaches to construct similariteis between species (see below), ee found that the most consistent results are obtained when similarity is determined through the Spearman correlation coefficient. In this case, the similarity of species $i$ and $j$ can be expressed as 
\eq{
S_{ij} =r_s(\vec{a^{(i)}}, \vec{a^ {(j)}}) = \frac{{\rm cov}(\vec{rg_i}, \vec{rg_j})}{\sigma_{i} \sigma_{j}}, \quad i\ne j 
}
where $\vec{rg_i}$ and $\vec{rg_j}$ are vectors of rank variables for the biomasses of species $i$ and $j$ in different samples, ${\rm cov}(\vec{rg_j}, \vec{rg_j})$ is the covariance between these variables, and $\sigma_{i}$ and $\sigma_{j}$ the standard deviations of the rank variables. 

As we are not interested in the similarity of a species to itself we set $S_{ii} = 0$. 
Moreover, as the correlation can be negative we rescale it as $S_{ij}\to (S_{ij}+1)/2$, 
which maps the values to the unit interval. 

An alternative approach (not used in the results in the maintext) is based on defining species dissimilarity as the distance between vectors of either normalized biomass values $\vec{b^{(i)}}=\vec{a^{(i)}}/\sigma_i$ or standardized biomass values, $\vec{b^{(i)}}=(\vec{a^{(i)}}-\mu_i)/\sigma_i$, where $\mu_i$ is the mean value and $\sigma_i$ the standard deviation of the biomass of species $i$ across all samples. 
We compute the dissimilarity $d_{ij}$ between species as the Euclidean distance between these vectors 
\eq{
  d_{ij} = \sqrt{\sum_{k} {\left(b_{(k)}^{(j)}-b_{(k)}^{(i)}  \right)^2}}
  }
and then define the elements of the similarity matrix, $S_{ij}$, as the inverse of the dissimilarity, such that 
\eq{
  S_{ij}=\frac{1}{d_{ij}}    
}
for $i\neq j$ and $S_{ij}=0$ for $i=j$.

\subsection{Thresholding} 
Long distance comparisons are unreliable and constitute a source of noise that can swamp the signal. Hence we want to eliminate all but the the largest entries in the similarity matrix. We do this examining each row of $\bf S$ and setting all but the 10 largest entries to zero. Thresholding more aggressively has a stronger denoising effect, but risks splitting the network into disconnected components. The value of 10 is often a good middle ground that has been used in several previous publications \cite{barter2019manifold,ghafourian2020wireless,ashkaanBacteria}. We confirmed that for the present dataset the links described by the similarity matrix form a spanning component. 

The thresholding procedure can leave us with an asymmetric matrix. For numerical reasons it is generally desirable to maintain the symmetry. We therefore resymmetrize the matrix by the operation
\eq{
S_{ij} \to {\rm max}(S_{ij},S_{ji}) 
}

\begin{figure}[ht!]
\centering
\includegraphics[width=0.8\textwidth]{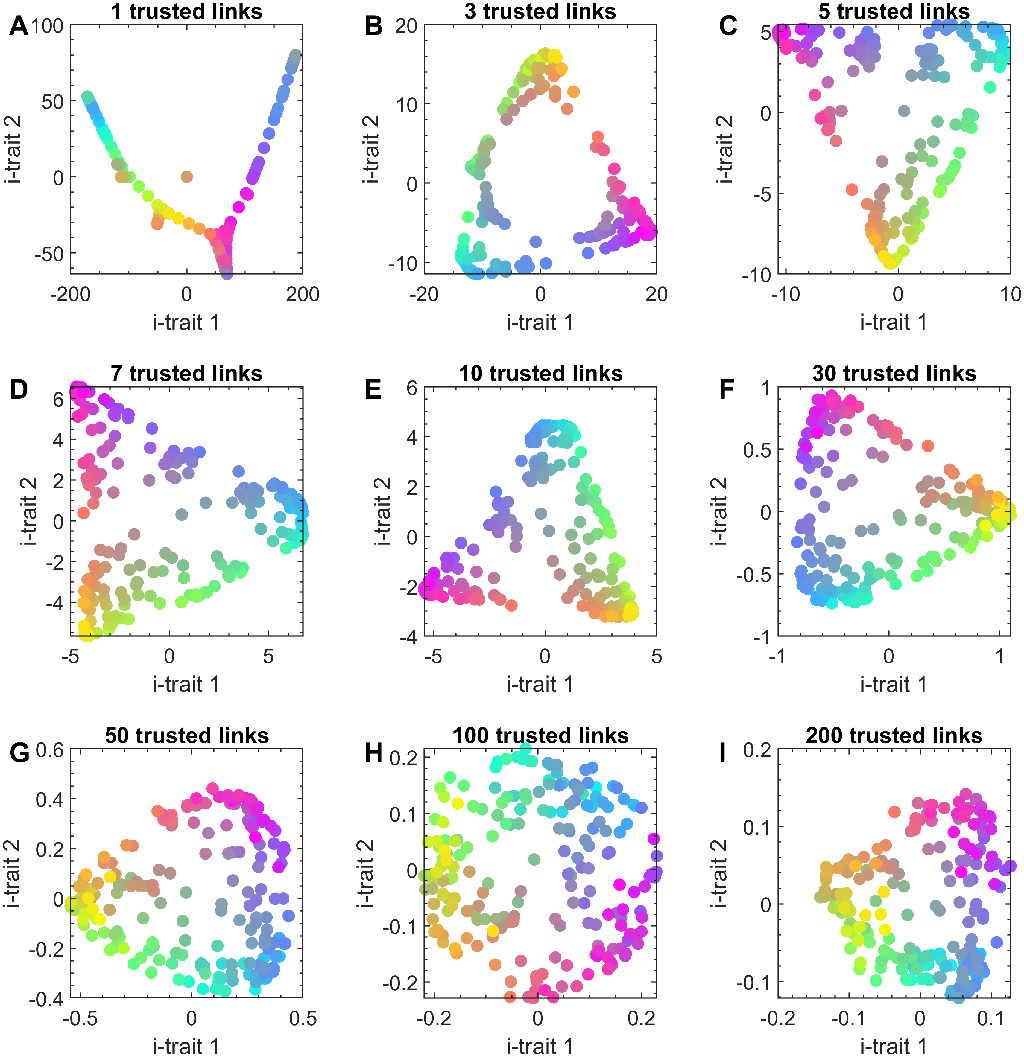}
\caption{\label{figThresholds}
Effects of the number of trusted links in the similarity matrix on the diffusion map. Shown is the i-trait space constructed from similarity matrices with different number of trusted links (see titles). Good results are obtained for a broad range of intermediate threshold values (ca.~5--30). If thresholds are chosen too small the network becomes fragmented. If thresholds are chosen too large the results become noisy. In the different panels the orientation of the reconstructed triangle changes due to a combination of the nature of eigenvectors and peculiarities of the synthetic dataset used in this demonstration (see text). 
}
\end{figure}

The effect of different choices of thresholds is illustrated in Fig.~\ref{figThresholds}. For the simulated dataset from the paper good results are obtained for a range of thresholds that lies roughly between 5 and 30. At the value of 30 the quality of the reconstruction starts to degrade due to the noise that is introduced by the long-range comparisons, which leads to a loss of shape information, particularly around the tips of the triangle. At 5 the network internally becomes too sparse leading to growing holes within the body of the triangle and eventual fragmentation. 

Note also that in the figure the orientation of the triangle changes from panel to panel. This is due to a combination of two effects. First, fundamentally, matrices specify their eigenvectors only up to a factor. Hence if $\vec{v}$ is an eigenvector with eigenvalues $\lambda$ (i.e. ${\bf L}\vec{v}=\lambda \vec{v}$) we can rescale the eigenvector by a factor $c$ and it will still be an eigenvector (${\bf L}(c\vec{v})= c{\bf L}\vec{v}=\lambda (c\vec{v})$). Hence by default a matrix only gives us the direction, but not the length of an eigenvector. While we impose a normalization condition such that the length of the vector is constrained to 1, this still allows for two different orientations of the vector. In diffusion-mapping we can use information contained in the eigenvalue to recover the best choice for the length of the eigenvector. However, note that $c$ can also be negative, which means the orientation of the eigenvector is still arbitrary and hence can flip between two configurations between trails. 

In our test data there is another effect that is due to the peculiarity of the simulated dataset. In the numerical experiments the trait space was chosen as a symmetric triangle, so there are three major axis of near identical length. Hence the first two eigenvectors are of exactly equal importance and their order can change from trial to trial due to numerical effects. 

\subsection{Laplacian} From $\bf S$ we construct the row normalized Laplacian $\bf L$ defined by 
\eq{
L_{ij}=\left\{\begin{array}{l l} -S_{ij}/\sum_j S_{ij} & \quad\quad i\neq j\\
1 & \quad\quad i=j  \end{array}\right. .
}
and solve the eigenvalue problem
$${\bf L} \vec{v_i} = \lambda_i \vec{v_i}, \qquad i=1...n.$$

\subsection{Eigenvectors and eigenvalues} 
All further analysis builds on the eigenvalues and eigenvectors of $\bf L$. We compute these eigenvalues and eigenvectors numerically using the function eig() from MATLAB 2021, which solves eigenvalue problems using the QR algorithm. 

The Laplacian always has at least one eigenvalue at zero. The multiplicity of this eigenvalue is identical to the number of components in the network. In our case, the network has only one component, hence the zero eigenvalue has multiplicity one. We denote this eigenvalue as $\lambda_0$. The corresponding eigenvector $\vec{v}_0$ contains no further information. All other eigenvalues are positive and the corresponding eigenvectors contain trait information. 

Since the Laplacian $\bf L$ is a symmetric matrix, all eigenvalues are real. The eigenvectors corresponding to small eigenvalues  are respectively more important, i.e.~explain more of the variation along the manifold. Hence the most important eigenvector $\vec{v_1}$ is the vector corresponding to the smallest non-zero eigenvalue. The eigenvector $\vec{v_2}$ corresponding to the second smallest non-zero eigenvector corresponds to the second most important trait and so on.  

Each eigenvector contains $n$ elements which assign a proxy trait value to each of the $n$ species. As the eigenvalues are inversely related to the importance of the trait we define the value of trait $k$ of species $i$ as $v_{k,i}/\lambda_k$.

\subsection{Diffusion distance}
Once the trait values have been computed the dissimilarity can be quantified by the distance in trait space. To compare two species we compute the euclidean distance between species in the i-trait space where the species traits are now given by the eigenvector elements corresponding to the species, rescaled by the respective eigenvalue. Hence the distance between two species $i,j$ is
\eq{
d_{ij}=\sqrt{\sum_k \left(\frac{v_{k,i}-v_{k,j}}{\lambda_k}\right)^2}
}


\section{Exploratory Analyses}
In this section we explain the calculation of functional diversity and present some supporting information on the analysis of the simulated data. Moreover, we present some results from exploratory analysis that lead up to the selection of the Spearman correlation coefficient as our primary similarity measure. 

\subsection{Functional diversity}
There are many ways to assess functional diversity. These methods have various advantages, and the choice of an appropriate index may depend on the specific problem~\cite{Legras2020FuncDiv}. In this article, we use the Rao’s quadratic entropy~\cite{BottaDukat2005Rao}, because this index is sensitive not only to the specific species traits, but also to variation in species abundances. Functional diversity calculated as the Rao index for sample $k$ equals
$$FD_k=\sum_{i=1}^{n-1}\sum_{j=i+1}^{n}{d_{ij} p^{(i)}_k p^{(j)}_k} \ ,$$ 
where $p^{(i)}_k=a^{(i)}_k/\sum_j a^{(j)}_k$ is the relative biomass of species $i$ in this sample. This index represents a weighted average functional distance between all pairs of species $i$ and $j$, where the weighting factor $p^{(i)} p^{(j)}$ is the probability that one of two randomly selected individuals belongs to species $i$ and the other to species $j$.

\begin{figure}
\includegraphics[width=\textwidth]{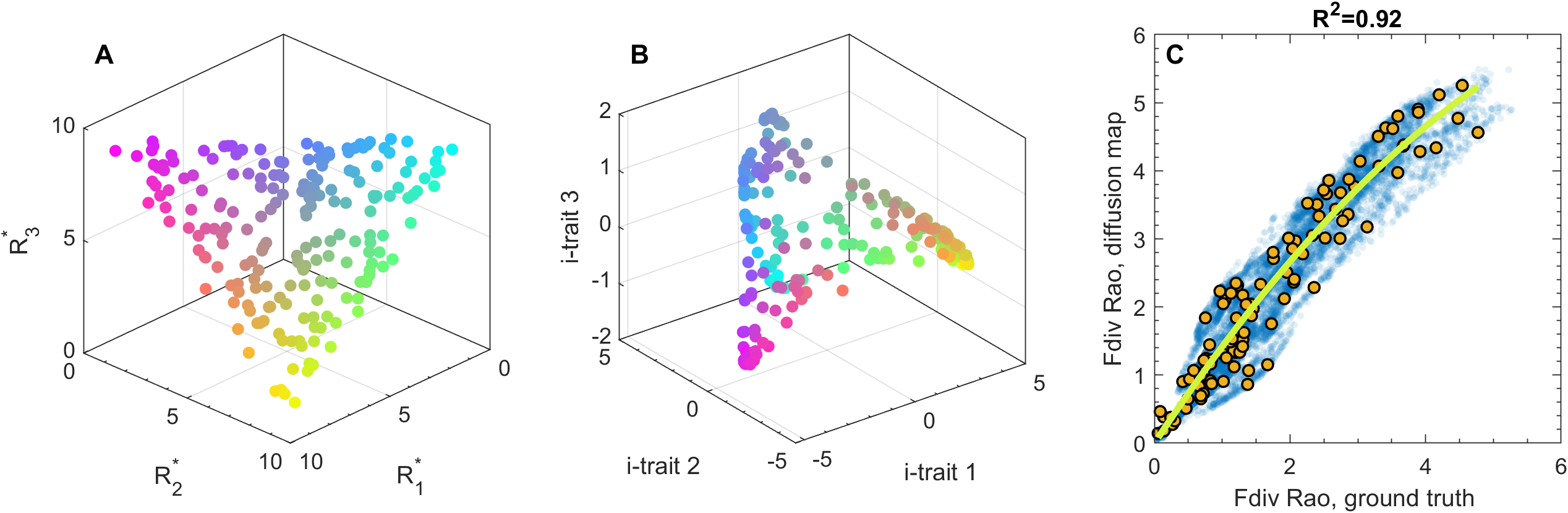}
\centering
\caption{Embedding using the third infered i-trait. Same as Fig.~3, but with a 3D representation of the species i-traits in B. Note the scale on the third axis. Even if more dimensions are used the inferred manifold remains a two-dimensional object to good approximation. 
but based on PCA of species similarity matrix calculated as Spearman correlation coefficients. 
}
    \label{fig3d}
\end{figure}

\subsection{Effect of data availability}
\begin{figure}[ht!]
\centering
\includegraphics[width=0.8\textwidth]{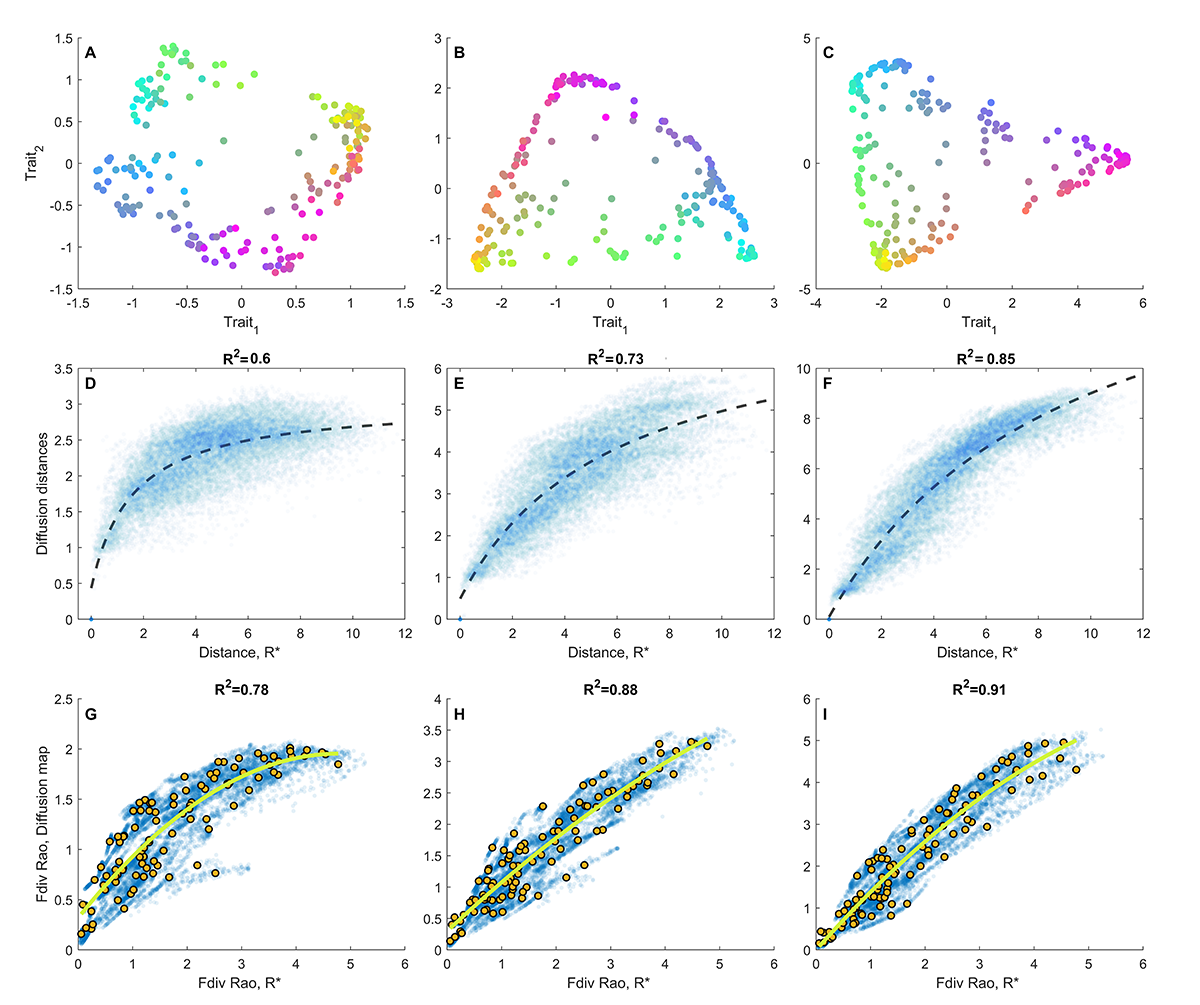}
\caption{Effects of data availability. Shown is the i-trait space (top row), a scatter plot of ground truth ($R^*$) trait distances vs.~i-traits distances (center row), and a scatter plot of estimated functional diversity vs.~ground truth functional diversity (bottom row) (\emph{cf.} Fig.~2). We compare different results using 100 (left column), 200 (center column), and 600 (right column) simulated grids to produce a diffusion map. The comparison and the indicated $R^2$ values in the second and third row illustrate how limited data availability degrades the inference result.}
 \label{fig:SSamlEff}
\end{figure}

In the main text we report the results of trait inference with simulated data from 800 grids. To explore how data availability affects the trait inference we repeat the inference using 100, 200, and 600 grids. For each of these trials we show the i-trait space and the estimated functional diversity (shown for the full data set in the main text), as well as distances between traits measured in terms of $R^*$ values vs.~diffusion distances (Fig.~\ref{fig:SSamlEff}). 

Using a small number of 100 grids the inference is very noisy, yielding a square rather than a triangular shaped trait space. The results of the inference improve steadily as we increase the number of grids used. With a number of 600 grids the quality of the inference is comparable to that of the 800-grid result shown in the main text.  

The diffusion map based on 100 grids allows us to reliably distinguish only species at small to intermediate functional distances in $R^*$ space, however inferred functional distance fluctuates around a constant level when the $R^*$ functional distance is large (Fig.~\ref{fig:SSamlEff}D, G). The correlation between both functional distances and functional diversities measured in $R^*$ space and in the i-trait space increase with the number of samples included, and the relationships approach to monotonically increasing reversible functions (Fig.~\ref{fig:SSamlEff}, middle and bottom row). This means that a sufficient input data volume provides a correct mapping for the entire range of functional distances.

\subsection{Selection of a distance measure}
When the original species traits are known, it is easy to estimate the amount of input data needed to construct a diffusion map with a targeted accuracy. For field data, however, we need an additional method to estimate the uncertainty for a given dataset and the possibility of reducing the  uncertainty by increasing the amount of data. To be able to achieve this we suggest to use bootstrapping of samples, a common machine learning method for estimating the rate of uncertainty reduction with the number of training samples.
This approach enables us to estimate the uncertainty of diffusion maps and allows the analysis of learning curves. 

We denote a similarity matrix constructed by using $m$ samples as ${\bf S}(m)$. To calculate this matrix we use bootstrapping, i.e., we select a random set of $m$ out of the $n$ samples with replacement. Thereby, on the basis of $n$ samples we obtain a reduced set of $m$ samples, which will probably include some samples several times, but have statistical properties close to the original set of samples. A bootstrap sample of $n$ out of $n$ samples will include, on average, 63\% of the original data, while 37\% will remain unused. This approach thus gives an estimate of the  uncertainty of the similarity matrix generated from the $n$ samples. 

To estimate the uncertainty, we quantify the 'difference' between a similarity matrix based on a subset of $m$ samples and a similarity matrix based the full set of $n$ available samples by the correlation between the elements of these matrices 
\eq{
V(m)={\rm cov}\left({\bf S}(m),{\bf S}(n)\right)
}

\begin{figure}
\includegraphics[width=\textwidth]{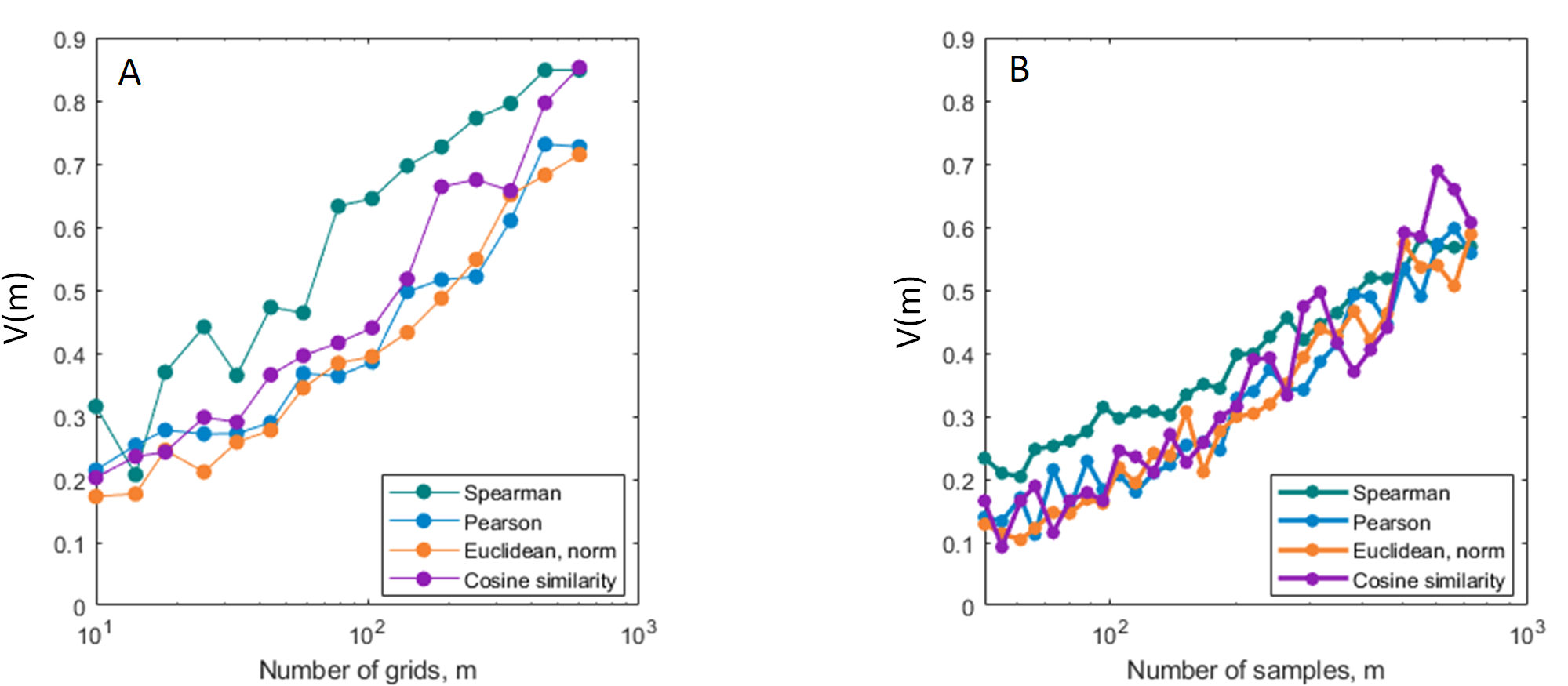}
\centering
\caption{Learning curve of the similarity matrix for different distance metrics. In both the simulated data (A) and Baltic Sea samples (B) the Spearman correlation provided the most robust indicator (higher values of $V(m)$) of species similarity over a wide range of values (see text for details).}
    \label{fig:SCorrOnM}
\end{figure}

Computing $V(m)$ over an increasing size $m$ of the subset yields a ``learning curve'' (Fig.~\ref{fig:SCorrOnM}).  Constructing such curves using different common notions of distance between species (see Supplement section 2.2) makes it possible to choose the metric that gives the most robust result.  
In this test the Spearman correlation performed very well, both for simulated and empirical data, yielding a good accuracy with limited data. Based on these results we choose the Spearman correlation as our primary notion of similarity between species. 

The value $V(n)$ gives an estimate of the uncertainty in ${\bf S}$ when all $n$  samples are used. Extrapolating these curves for $m>n$ provides a rough idea of the number of samples needed to calculate the similarity matrix with the desired accuracy. For the simulated data, when all samples are used, the correlations achieve values from 0.7 to 0.85 depending on the distance metric (Fig.~\ref{fig:SCorrOnM}A), and the correlation continues to grow monotonically. To obtain a substantial reduction of uncertainty in this case one need at least to double the number of samples.  In the field data, compared to the model data, the number of independent variables (species) is larger and the number of observations is smaller. This increases the uncertainty of the similarity matrix, and the correlations obtained for the complete dataset achieve maximal values of $V(m)=0.6$ (Fig.~\ref{fig:SCorrOnM}B). Thus, an increase of the monitoring area, frequency, or time-span should be able to significantly improve the accuracy of both the similarity matrix and i-traits. 

\subsection{Comparison of diffusion maps and PCA}
Principal Component Analysis (PCA) is a very common data analysis procedure that identifies directions of large variation in a data cloud. Hence PCA can be used as a simple manifold learning method if the true manifolds in the data are close to linear. However, as we now demonstrate, using PCA for i-trait results in reduced accuracy. 

\begin{figure}
\includegraphics[width=\textwidth]{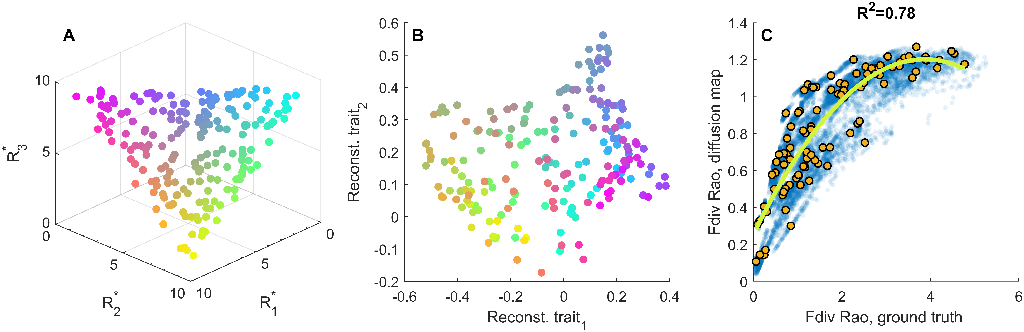}
\centering
\caption{Poor performance of PCA. The same as Fig.~2
but based on PCA of species similarity matrix calculated as Spearman correlation coefficients. 
}
    \label{fig:STraitsPCA}
\end{figure}

\begin{figure}
\includegraphics[width=\textwidth]{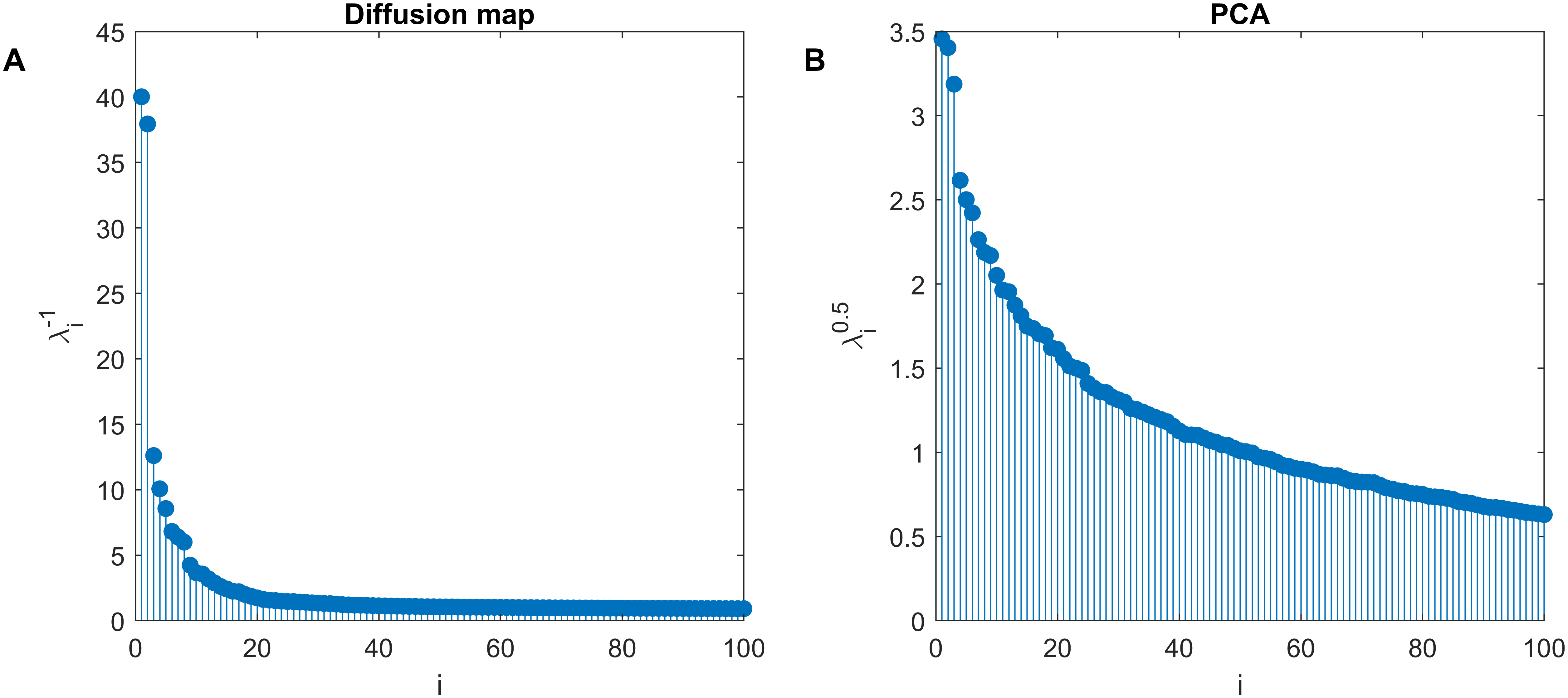}
\centering
\caption{Comparison of transformed eigenvalues (loadings) calculated for diffusion maps (A) and PCA (B). The transformed eigenvalue shows the variability in data described by the given axes in the eigenvector space. As show, the representation of data in diffusion map space is much more condense in lower dimension of the eigenvector space, allowing us to use only a few first components to characterize the data. 
}
    \label{fig:SDMPCAEigs}
\end{figure}

In the analysis of model data significant advantages of the diffusion map over PCA become apparent. We can highlight three main differences. First, the shape of trait distribution in the diffusion map, compared to PCA, is closer to the original distribution shape (cf.~Fig.~2A,B and Fig.\ref{fig:STraitsPCA}). Second, the representation of data by diffusion map compared to PCA is much more condense in lower dimensions (Fig.~\ref{fig:SDMPCAEigs}). Finally the fractal dimension of the trait manifold infered by the diffusion map is closer to the dimension of the original manifold than that obtained by PCA (Fig.~\ref{fig:SFractalDim}).

\subsection{Mapping dynamical data}
\begin{figure}[ht!]
\centering
    \includegraphics[width=0.8\textwidth]{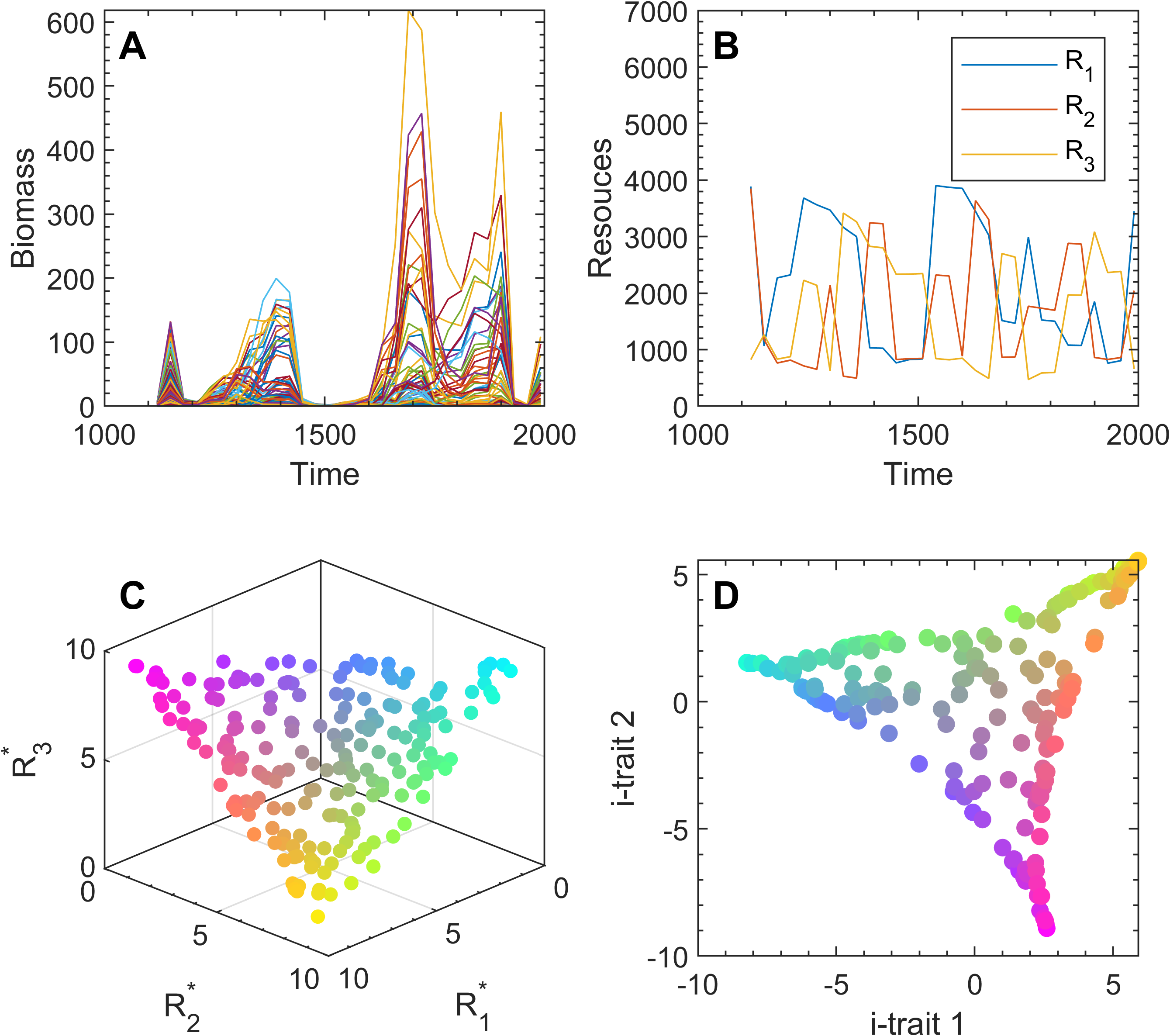}
    \caption{Trait inference for a dynamical model. Times series of species (A) and resources (B) remain dynamic in simulations where the resource distribution was updated following the same random process every 50 time units. Grid-averaged species biomass and resource concentrations. (C) Species resource-requirement ($R^*$). Color indicates the resource ratio preferred by a species. (D) Species i-traits generated by diffusion mapping simulated biomass data. Colors are the same as in C, illustrating that neighborhood relationships are mostly inferred correctly. }
    \label{figDynamic}
\end{figure}

We also explored whether the stronger ongoing dynamics observed in real data would impinge our ability to construct i-traits. The case can be made that non-stationary dynamics help-rather than hinder the inference as they provide us with more independent information we therefore ran a smaller test using only 100 grid replicas, where resources were reshuffled after 50 time units. Already based on this limited amount of data a comparatively good inference of the trait space was possible (Fig.~\ref{figDynamic})

\subsection{Dimensionality of the data space}
To compare the representation of the original trait manifold by a minimal set of i-traits obtained by PCA and diffusion map, it is convenient to use the obtained eigenvalues. For diffusion maps, species traits are defined as $t_{j, i} =v_{j,i}/\lambda_j $, while PCA ordination of species uses the so-called scaling 2 representation where species traits are defined as $t_{j, i} = \sqrt{\lambda_j }v_{j,i} $ \cite{legendre2012numerical}. Since the absolute values of the eigenvectors $\vec{ v_{j} }$ equal one, the comparison of the ranges of species trait variation is reduced to a comparison of the transformed eigenvalues: $1/\lambda_j $ for the diffusion map and $\sqrt{\lambda_j }$ for PCA (Fig.~\ref{fig:SDMPCAEigs}). Note that $\sqrt{\lambda_j }$ is termed as loadings in PCA analysis. 

The initial manifold of species traits is a two-dimensional object (Fig.~\ref{fig:STraitsPCA}A), and the values of first two inverted eigenvalues obtained by diffusion mapping significantly exceed the subsequent eigenvalues (Fig.~\ref{fig:SDMPCAEigs}A), while for PCA this difference is much smaller and even traits defined by eigenvectors with an index greater than 50 carry a significant part of information (Fig.~\ref{fig:SDMPCAEigs}B). Thus, diffusion maps compared to PCA are much better at concentrating information about species traits in a lower-dimensional space. 

\begin{figure}
\includegraphics[width=8cm]{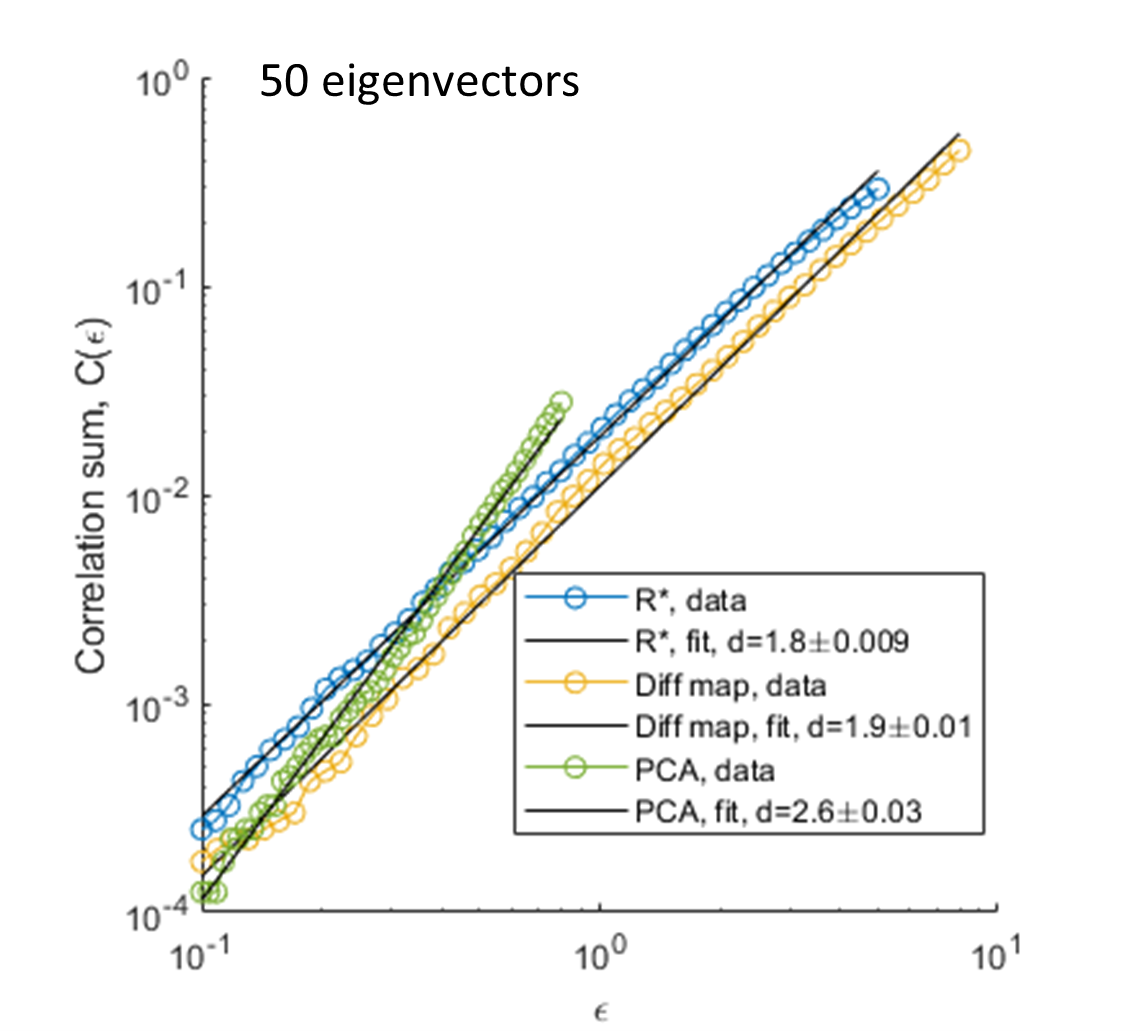}
\centering
\caption{Estimation of the correlation fractal dimension $d$. The estimated dimension of the original trait distribution in space $R^*$ is $d=1.8$, which is close to the actual dimension $d=2$, since we put all $R^*$ on a plane. The dimensionality of the surface in the i-trait space from the diffusion map is $d=1.9$. The distribution in PCA space has a larger dimensionality, $d=2.6$, than the actual distribution, which implies that the data points in PCA space are likely located with random displacement around a two-dimensional surface, effectively creating an object with a fractional dimension greater than 2. The fractal dimension was estimated in the space of the first 50 eigenvectors.
}
    \label{fig:SFractalDim}
\end{figure}

\begin{figure}[t!]
\includegraphics[width=\textwidth]{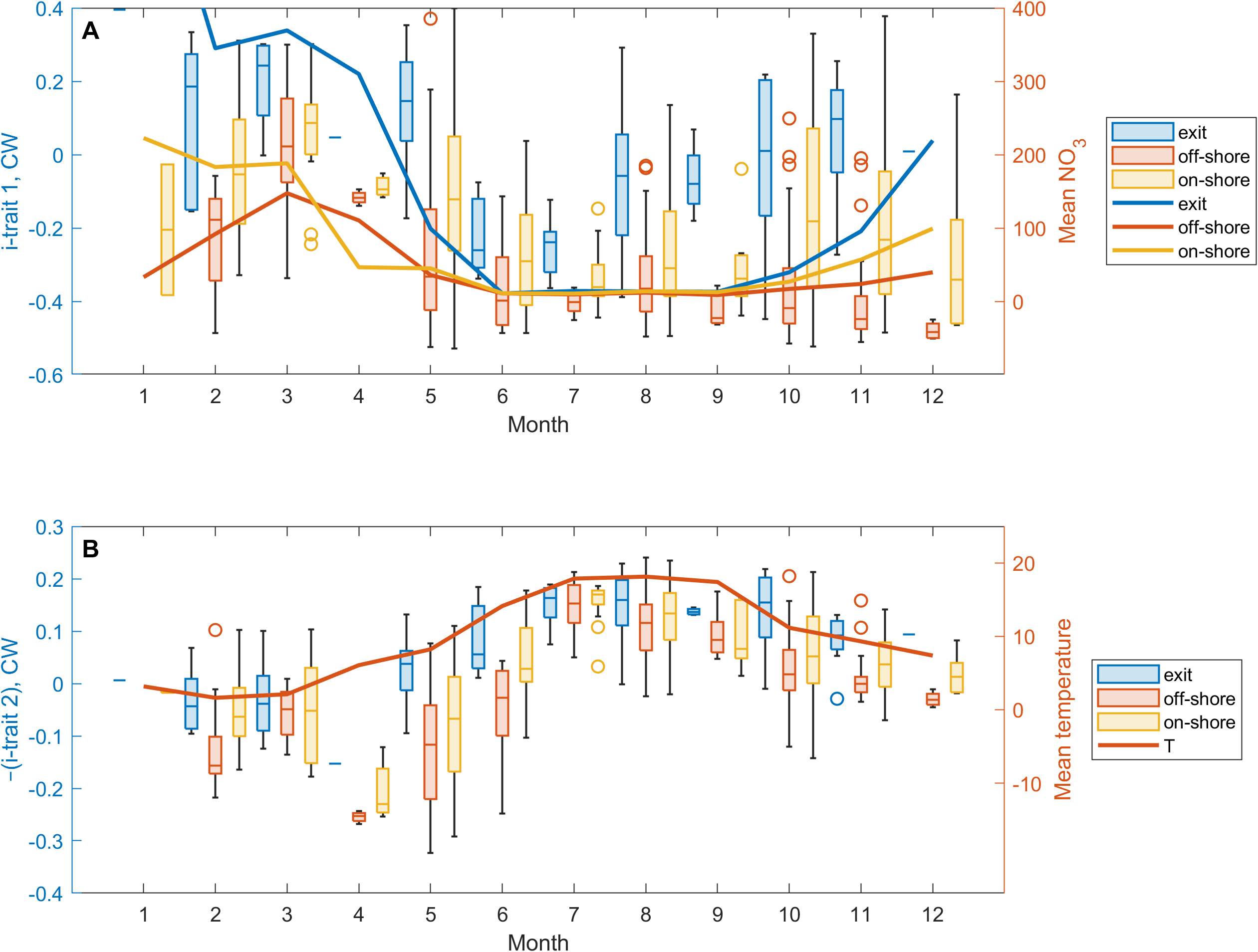}
\centering
\caption{Seasonal trends in environmental conditions and corresponding community-weighted i-traits. The box-plot shows the distribution of community weighted mean of i-trait 1 (nitrogen) and i-trait 2 (temperature) for different months and station locations (off shore, on shore, and the Curonian Lagoon exit). Solid lines show the average of NO3 (grouped by locations) and temperature. (A) The variation of i-trait 1 (nitrogen) reaches the highest values in spring and drops to a minimum in summer, corresponding to the period of summer nutrient depletion. Autumn and winter values of i-trait 1 depend on station location, the lowest values being observed for off-shore stations, while species correlating with high nitrogen levels are present at the Curonian Lagoon exit. (B) The variation of the community weighted mean of i-trait 2 is in good agreement with the water temperature, which was almost the same across all locations. As i-trait 2 and temperature are negatively correlated, here we reversed the sign of i-trait 2 to make the comparison clearer. Note that the low i-trait 2 values in April should be considered as an outlier, because the sample in April was taken only in 2000. }
    \label{figTraitDyn}
\end{figure}

The inferred manifold is a multidimensional object, with the number of dimensions equal to the number of species. However, we expect that the resulting distribution of species traits should keep the original distribution dimensionality and be located on a two-dimensional surface. To estimate the dimensionality of the resulting distribution, we calculated its correlation fractal dimension in the space defined by the first 50 eigenvectors. The fractal dimension of the original distribution, $d=1.9$, is less than 2, because a random distribution is never perfectly uniform and contains some halls. The diffusion map fractal dimension $d=1.8$ is close to the original value, but the PCA distribution dimension exceeds it and equals 2.6  (Fig.~\ref{fig:SFractalDim}). Implying that the distribution obtained with PCA is closer to a 3D object than a 2D surface. 

\section{Analysis of the Baltic Sea data}
This section contains some additional notes on the Baltic Sea phytoplankton dataset and it's analysis. We use data from at 10 stations the Lithuanian coastal area of Baltic Sea which were regularly visited for 23 years (1993-2015). The data includes 730 samples of biomass of 516 species and environmental data, such as temperature, NO$_3$, NO$_2$, PO$_4$, pH, and salinity. 

\subsection{Adaptation to environmental factors}
Environmental data were measured at different depths for each sample. To obtain a single value for each sample, we interpolated these measurements over a range of depths from 1 to 10 m with resolution of 1 m and calculated the average value in this range. 

\begin{figure}
\centering
\includegraphics[width=0.75\textwidth]{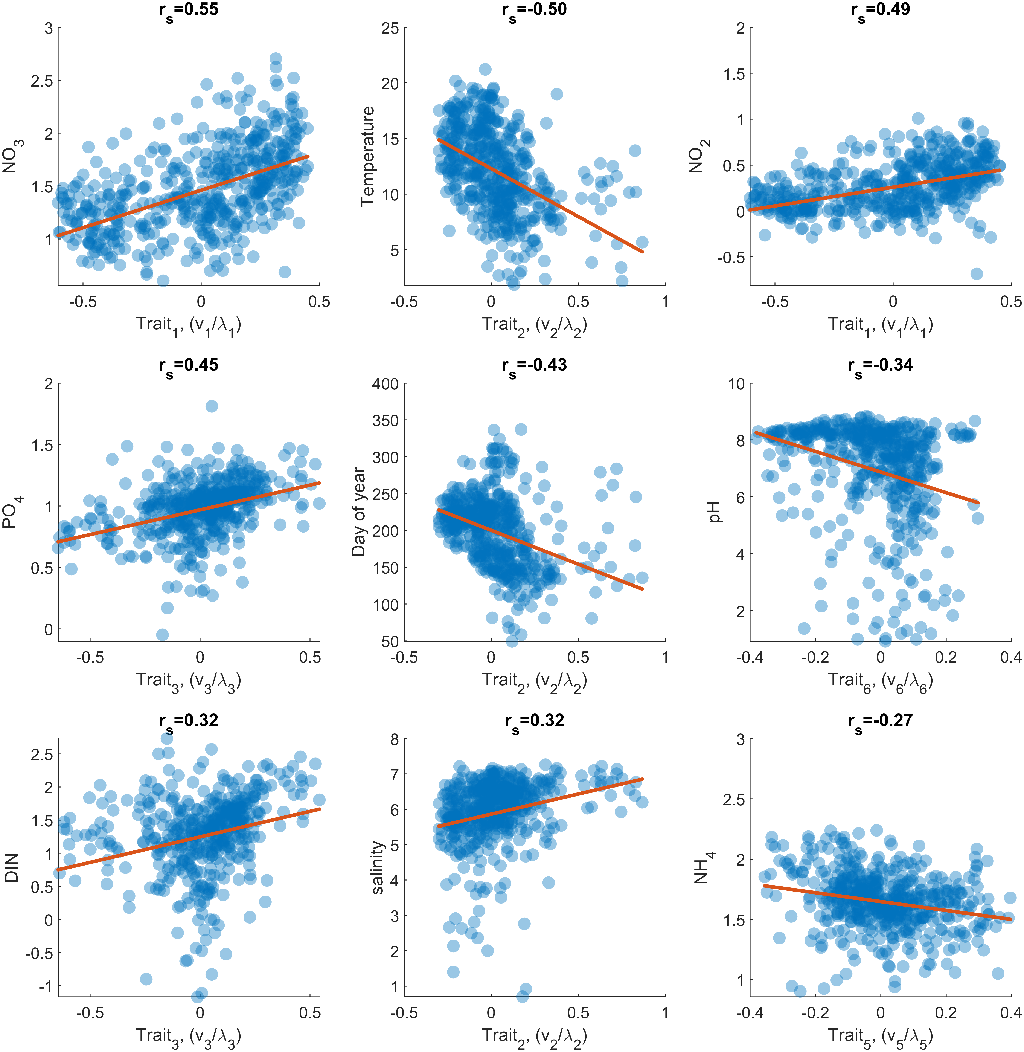}
\caption{Correlations between i-traits and species-specific environmental values. The top nine trait-environment pairs with the highest Spearman correlation coefficient are shown (see text for details). }
    \label{fig:STraitEnv}
\end{figure}

\begin{figure}
\includegraphics[width=\textwidth]{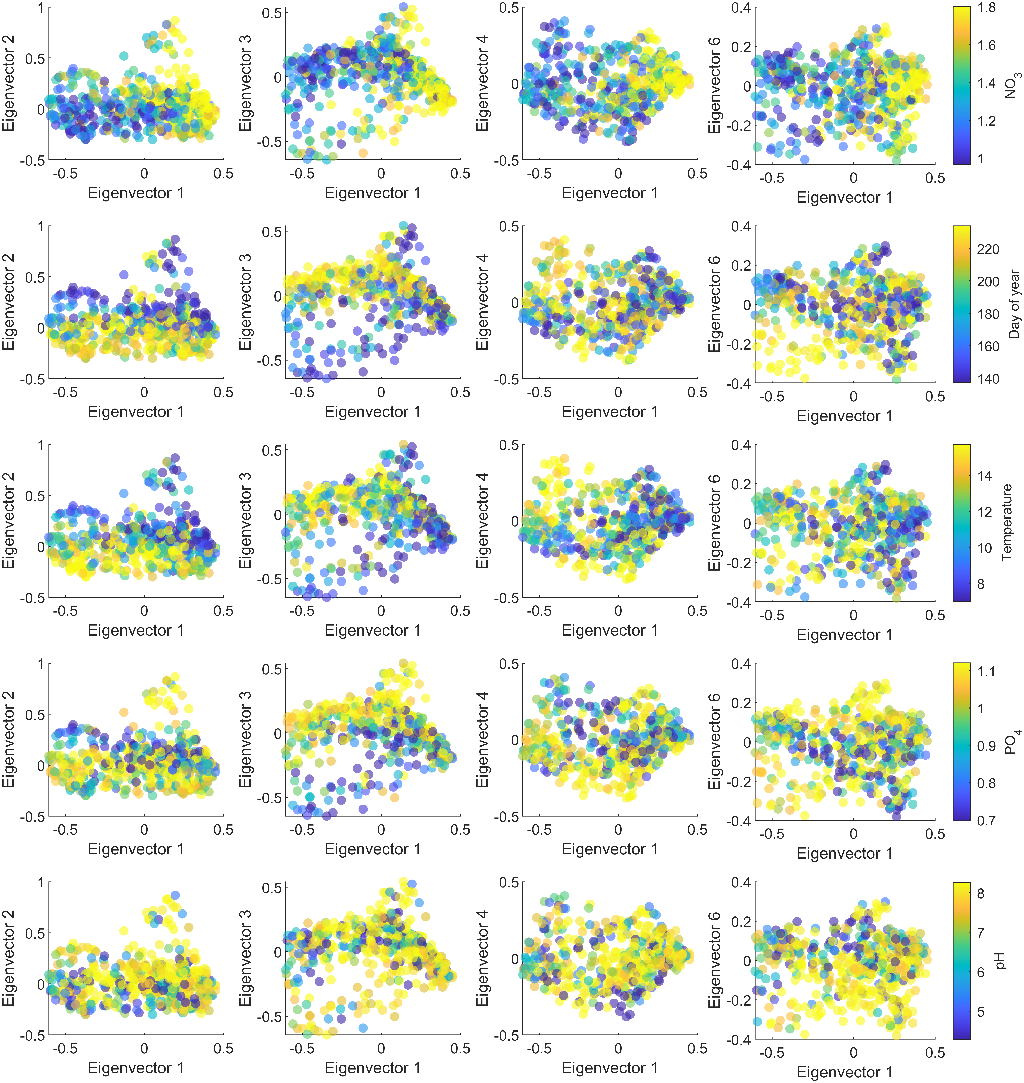}
\centering
\caption{Projection of diffusion map on different axes. Color shows the value of the species-specific environmental factor.}
    \label{fig:SDiffMaps}
\end{figure}

We estimated species specific environmental condition as the average environmental condition weighted with species biomass 
\[\hat{E} ^{(r,i)}=\frac{ \sum_{j=1}^n a_j^{(i)} E_j^{(r)}  }{\sum_{j=1}^n a_j^{(i)}}\]
where \(a_j^{(i)}\) is the biovolume of species \(i\)  in sample \(j\), \(E_j^{(r)}\) is the environmental factor \(r\) in this sample, and $n$ is the number of samples. In this way we obtain the species-specific day of year, temperature, nutrients, pH, etc., which are used to colorize the diffusion maps (Fig.~3 and Fig.~\ref{fig:SDiffMaps}), and to identify species trait-environment pairs (Fig.~\ref{fig:STraitEnv}).

To find the best matching between traits and environmental factors shown in Fig.~3 in the main text, we performed a cross-correlation analysis of the relationships between the environmental factors and the i-traits. We calculated the Spearman correlation between the measured environmental factors and the first ten proxy traits, selected then the trait-environment pair with the highest correlation for each environmental factor, and sorted the results by correlation in descending order. Fig.~\ref{fig:STraitEnv} shows the top nine trait-environment pairs. The first trait correlates with NO$_3$ ($r_S = 0.55$) and NO$_2$ ($r_S =0.49$); the second trait negatively correlates with temperature ($r_S = -0.5$), day of year ($r_S = -0.43$) and positively correlates with salinity ($r_S = 0.32$); the third trait correlates with PO$_4$ concentration ($r_S = 0.45$) and DIN ($r_S = 0.32$); the fifth trait correlates with NH$_4$ ($r_S = -0.27$), and the sixth trait with pH ($r_S = -0.34$). We found no strong correlation between the fourth trait and any of the environmental conditions, but this does not mean that this trait is not important, as it may reflect adaptation to factors missed in our data, such as zooplankton abundance, light radiation, or water turbidity. 

By projecting the distribution of species traits from the multidimensional diffusion map space onto different species trait axes, we obtain more details about the relationships between environmental factors and proxy traits. This is shown in Fig.~\ref{fig:SDiffMaps} which is an extended version of Fig.~3 from the main text.  As shown, adaptation to NO$_3$ correlates only with the first proxy trait, and adaptation to temperature correlates only with the second trait. At the same time, the day of the year is described by the second, third, and sixth trait (with decreasing correlation). This can be explained by the fact that this parameter strongly positively correlates with temperature, however species dominant in spring and autumn may have the same optimal temperature, but be characterized by different days of the year. Adaptation to PO$_4$ concentration is also related to several traits, but only the relationship with the third trait is monotonic, while the relationships with the second, fourth, and sixth trait is unimodal.

\begin{figure}[ht!]
 \includegraphics[width=0.95\textwidth]{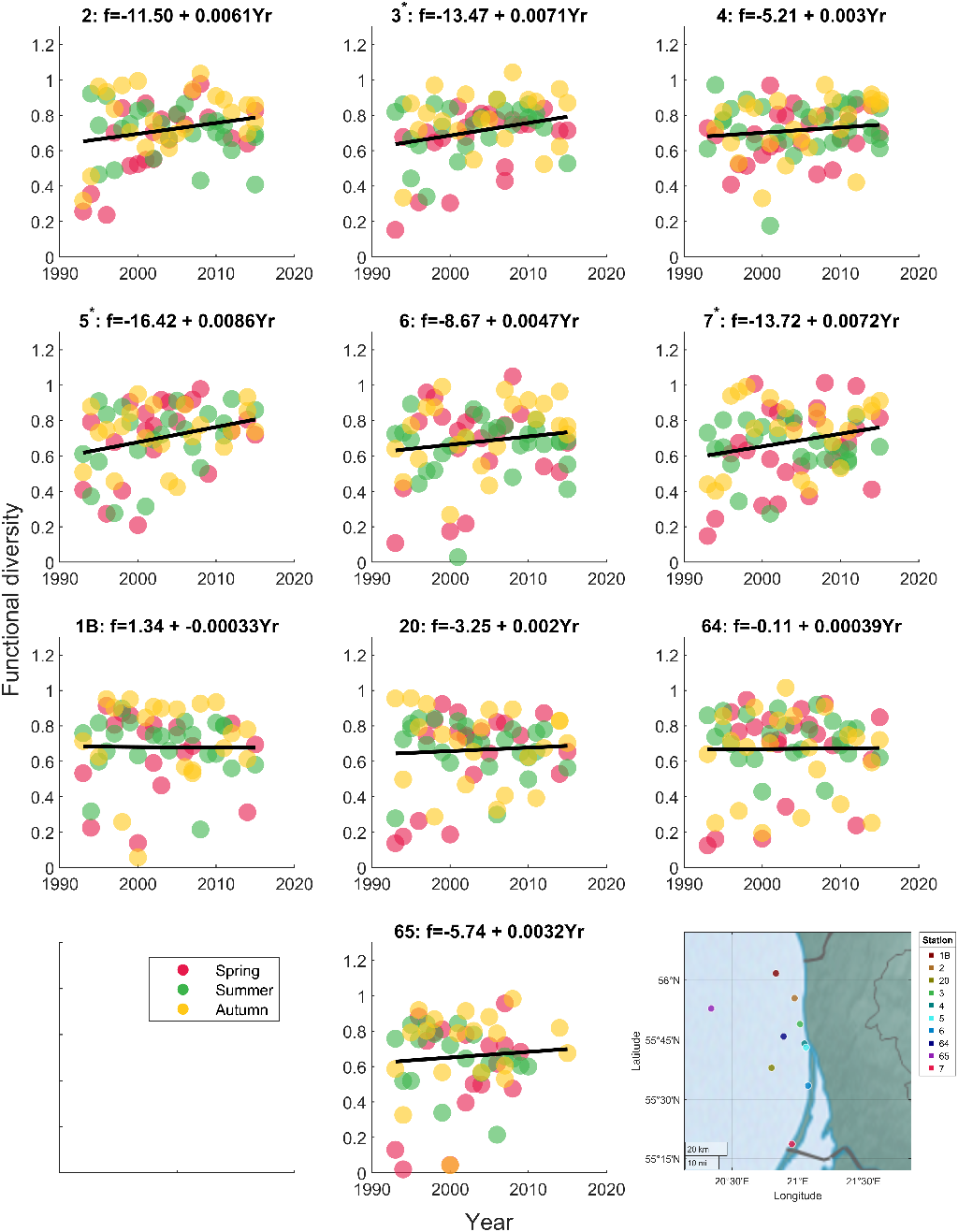}  
    \caption{Temporal trends of functional diversity at different stations. Functional diversity at coastal stations 2, 3, 5, 6, 7 has increased over the time of observations. This increase was significant ($p<0.05$) at stations 3, 5, and 7, marked with a star in the plot title. At coastal station 4 it was high during the entire monitoring period. In recent years, samples from all coastal stations exhibit only high functional diversity. By contrast, functional diversity at off-shore stations 1B, 20, 64, and 65 has a weak positive trend and varied over a wide range throughout the observation period.}
    \label{SStatTrends}
\end{figure}
 \begin{figure}[t!]
 \includegraphics[width=0.91\textwidth]{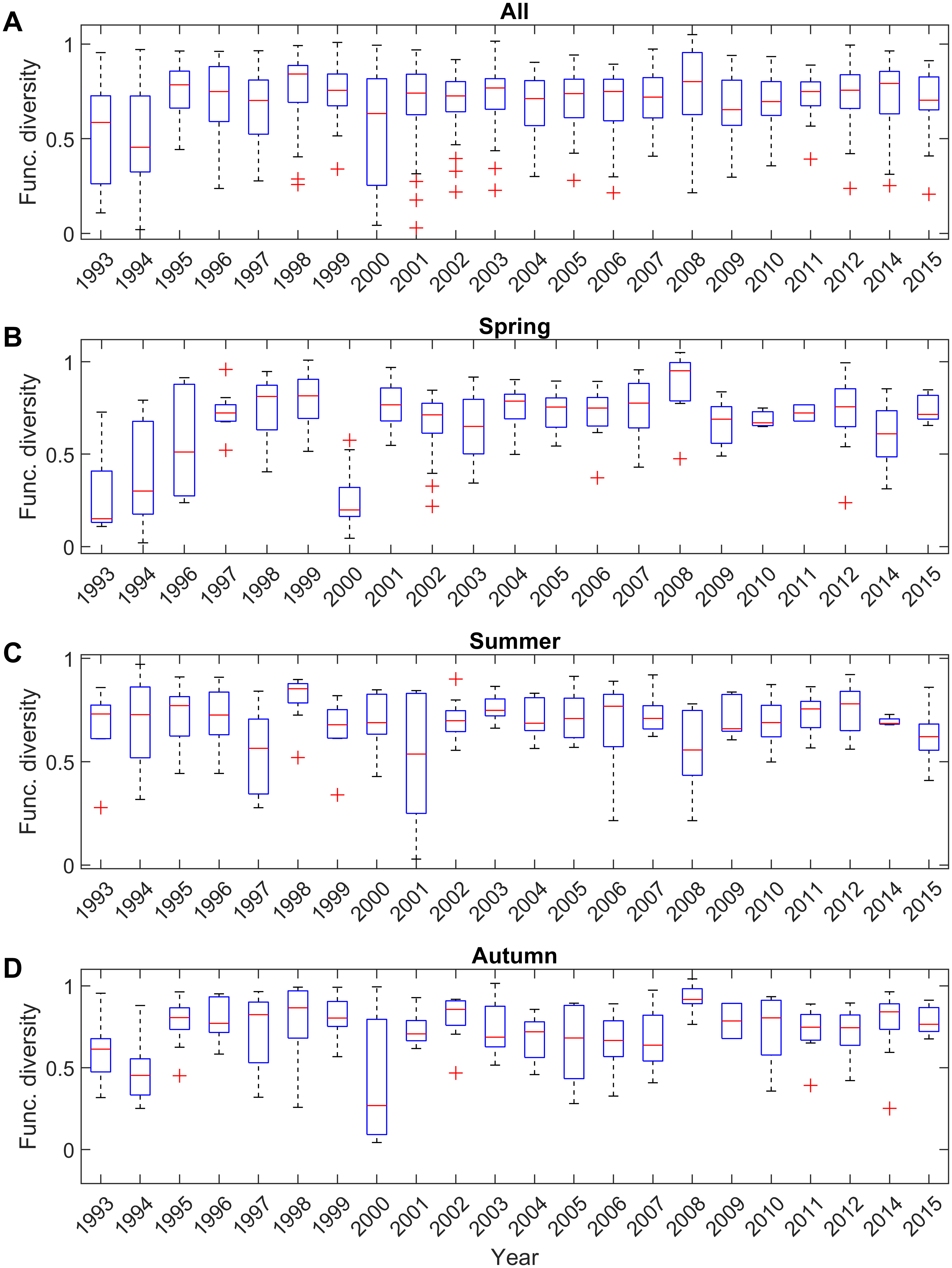}  
    \caption{Temporal trends of functional diversity at different seasons. Boxplots showing the variation and median values of functional diversity in dependence of the year. The variation in diversity across observations and stations, shown both for the whole year (top row) and separately for each season. The boxplots show the median values (red line), the bottom and top of each box are the 25th and 75th percentiles of the sample, and the whisker length is $\pm 2.7 \sigma$  (corresponding to 99.3\% for normally distributed data).}
    \centering
    \label{STrendsSeasons}
\end{figure}
 
 Fig.~\ref{STrendsSeasons} shows the dynamics of functional diversity grouped by year (upper row) and additionally by season (lower rows). Functional diversity has remained high on average, with the exception of 1993, 1994, and 2000 (Fig.~\ref{STrendsSeasons}A). Spring samples showed an increase in functional diversity early in the observation period (1993-1999), followed by a decline in functional diversity in 2000, after which functional diversity remained at consistently high levels (Fig.~\ref{STrendsSeasons}B). Extremely large variation between stations, characterized by the size of boxes in Fig.~\ref{STrendsSeasons}, was observed in the spring samples of 1993, 1994, and 1996, in the summer samples of 1994, 1997, and 2001, and also in the fall sample of 2000. 
 
\subsection{Dynamics of i-traits and functional diversity}
For comparison of the dynamics of functional diversity at different stations (Fig.~4), we carried out a regression analysis of the dependence of functional diversity on the year at different stations (Fig.~\ref{SStatTrends}). Since we did not identify a statistically significant seasonal effect on functional diversity, the regression analysis was performed without grouping the data by seasons, but the data were grouped by seasons for clarity when plotting the figure. All coastal stations, with the exception of station 4, show a positive temporal trend in functional diversity, associated with the fact that samples at the beginning of the observations show a wide range of diversity, while samples closer to the end of the observations are characterized mostly by high functional diversity. Coastal station 4, located at the exit of the Curonian Lagoon, is the only coastal station that shows only a small statistically insignificant temporal trend, which is explained by the fact that the functional diversity at this station was at a high level throughout the entire observation period. In contrast to the coastal stations, the samples from off-shore stations (1B, 20, 64, 65) do not show a positive trend in functional diversity and are characterized by a reduced level of functional diversity throughout the entire observation period compared to the coastal stations. 
The dynamics of some selected i-traits are shown in Fig.~\ref{figTraitDyn}


\end{document}